\documentclass{elsart}

\usepackage{graphicx}
\usepackage{amssymb}

\newcommand{\KK}{\mbox{\boldmath$K$}}

\newcommand{\MM}{\mbox{\boldmath$M$}}

\newcommand{\rr}{\mbox{\boldmath$r$}}

\newcommand{\QQ}{\mbox{\boldmath$Q$}}

\newcommand{\VV}{\mbox{\boldmath$V$}}
\newcommand{\vv}{\mbox{\boldmath$v$}}

\newcommand{\xx}{\mbox{\boldmath$x$}}

\begin{document}

\begin{frontmatter}

\title{\textbf{Uniqueness of thermodynamic projector and kinetic basis of molecular individualism}}

\author{Alexander N.\ Gorban \corauthref{cor1}}
\ead{agorban@mat.ethz.ch} \corauth[cor1]{Corresponding author: Department of Materials, Institute of
Polymers, Polymer Physics, ETH-Zentrum, Sonneggstrasse 3, ML J 27, CH-8092 Z\"urich, Switzerland.}
\address{Swiss Federal Institute of Technology, Z\"urich, Switzerland
\\ Institute of Computational Modeling RAS, Krasnoyarsk, Russia
\\  Institut des Hautes Etudes Scientifiques, Bures-sur-Yvette, France}

\author{Iliya V.\ Karlin}
 \ead{ikarlin@mat.ethz.ch}
\address{ Swiss Federal Institute of Technology, Z\"urich, Switzerland
 \\ Institute of Computational Modeling RAS, Krasnoyarsk, Russia}


\maketitle

\begin{abstract}

Three results are presented: First, we solve the problem of persistence of dissipation for reduction
of kinetic models. Kinetic equations with thermodynamic Lyapunov functions are studied. Uniqueness of
the {\it thermodynamic projector} is proven: There exists only one projector which transforms any
vector field equipped with the given Lyapunov function into a vector field with the same Lyapunov
function for a given ansatz manifold which is not tangent to the Lyapunov function levels.

Second, we use the thermodynamic projector for developing the short memory approximation and
coarse-graining for general nonlinear dynamic systems. We prove that in this approximation the
entropy production increases. ({\it The theorem about entropy overproduction.})

In example, we apply the thermodynamic projector to derivation the equations of reduced kinetics for
the Fokker-Planck equation. A new class of closures is developed, the kinetic multipeak polyhedra.
Distributions of this type are expected in  kinetic models with multidimensional instability as
universally, as the Gaussian distribution appears for stable systems. The number of possible
relatively stable states of a nonequilibrium system grows as $2^m$, and the number of macroscopic
parameters is in order $mn$, where $n$ is the dimension of configuration space, and $m$ is the number
of independent unstable directions in this space. The elaborated class of closures and equations
pretends to describe the effects of ``molecular individualism". This is the third result.

\end{abstract}

\end{frontmatter}





\section*{Introduction}

Reduction of description for dissipative kinetics assumes (explicitly or implicitly) the following
picture: There exists a manifold of slow motions in the space of distributions. From the initial
conditions the system goes quickly in a small neighborhood of the manifold, and after that moves
slowly along it.

There are three basic problems in the model reduction:
\begin{enumerate}
\item{How to {\bf construct} the slow manifold;}
\item{How to {\bf project} the initial equation onto the constructed slow manifold, i.e. how to split motions
into fast and slow;}
\item{How to {\bf improve} the constructed manifold and the projector in order to make the manifold more invariant and
the motion along it slower.}
\end{enumerate}

The first problem is often named ``the closure problem", and its solution is the closure assumption;
the second problem is ``the projection problem". Sometimes these problems are discussed and solved
simultaneously (for example, for the quasiequilibrium, or, which is the same, for MaxEnt closure
assumptions \cite{Janes1,Zubarev,KoRoz,Ko,Kark}). Sometimes solution of the projection problem after
construction of ansatz takes a long time. The known case of such a problem gives us the
Tamm--Mott-Smith approximation in the theory of shock waves (see, for example, \cite{GK1}). However
if one has constructed the closure assumption which is at the same time the {\it invariant manifold}
\cite{GK1,GKTTSP94,Lam}, then the projection problem disappears, because the vector field is always
tangent to the invariant manifold.

Let us discuss the initial kinetic equation as an abstract ordinary differential
equation\footnote{Many of partial differential kinetic  equations or integro-differential kinetic
equations with suitable boundary conditions (or conditions at infinity) can be discussed as abstract
ordinary differential equation in appropriate space of functions. The corresponding {\it semigroup}
of shifts in time can be considered too. For example, the Fokker-Planck equation in a potential well
$U(q)$ with a condition $U(q)/\|q\|^{\alpha}\rightarrow \infty$ for $\|q\|\rightarrow \infty$ and
some $\alpha
> 0$ generates an analytical semigroup. It allows to discuss the Fokker-Planck equation in such a
well on the same way as an ordinary differential equation. Sometimes, when an essential theorem of
existence and uniqueness of solution  is not proven, it is possible to discuss a corresponding shift
in time with the support of physical sense: the shift in time for physical system should exist.
Benefits from the latter approach are obvious as well as its risk.},
\begin{equation} \label{sys}
{d\Psi \over dt}= J(\Psi),
\end{equation}
where $\Psi=\Psi(q)$ is the distribution function, $q$ is the point in  configuration space (for the
Fokker-Planck equation) or in phase space (for the Liouville equation).

Let the closure assumption be given:
\begin{equation} \label{clas}
\Psi=\Psi(M|q),
\end{equation}
where $M$ is the set of macroscopic variables, which are coordinates on the manifold (\ref{clas}).
The tangent space $T_{M_0}$ for the manifold (\ref{clas}) in the point $M_0$ is the image of the
differential:
\begin{equation} \label{tanspa}
T_{M_0}=\mbox{im}(D_{M}(\Psi(M|q))_{M_0}.
\end{equation}
How to construct the dynamic equation for the variables $M$? This is the projection problem. The
equivalent setting is: how to project $J(\Psi(M_0|q))$ onto $T_{M_0}$? If $dM/dt=F(M)$ is the
equation for $M$, then the equation on the manifold is $d \Psi (M|q)/dt=(D_{M} \Psi (M|q)) \cdot
F(M).$

There exist three common ways to construct the projector onto $T_{M_0}$:
\begin{enumerate}
\item{Moment parametrization;}
\item{Spectral projectors of Jacobians for equation (\ref{sys})};
\item{Spectral projectors of ``symmetric part" of Jacobians for this equation.}
\end{enumerate}

The moment parametrization is the best way to ``hide" the projector problem in a natural way: Let the
macroscopic variables be defined not only on the manifold $\Psi(M|q)$, but in the neighborhood of
this manifold: $M=m(\Psi)$, with the identity $m(\Psi(M|q))\equiv M$. Then we can define $dM/dt$ in a
natural way:
\begin{equation} \label{momeq}
{dM \over dt} = (D_{\Psi}m(\Psi(M|q)))J(\Psi(M|q)).
\end{equation}
As it will be demonstrated below, this simple formula is appropriate only for the quasiequilibrium
(MaxEnt) approximation, because in other cases it leads to entropy decreasing for some initial
conditions and, hence, to a perpetuum mobile of the second kind (this happens in reduced equations,
of course, and not in reality).

The idea of slow-fast decomposition through spectral decomposition of Jacobian seems attractive (see,
for example, the theory of the so-called intrinsic low-dimensional manifold (ILDM) \cite{Maas}): Let
the spectrum of $D_{\Psi}J(\Psi)$ can be separated in two parts: $\mbox{Re}\lambda_{\rm sl} < A \ll B
< \mbox{Re}\lambda_{\rm fst} < 0$. There are two invariant subspaces which correspond to slow
($E_{\rm sl}$) and to fast ($E_{\rm fst}$) points of the spectrum. The suggested solution of the
projection problem is: The tangent space $T_M$ of the slow manifold should be not very different from
the slow invariant subspace $E_{\rm sl}$, and the projection of $J$ onto $T_M$ should be done
parallel to the fast invariant subspace $E_{\rm fst}$.

The eigenvectors and eigenprojectors of the non-selfadjoint operators may be very unstable in
calculations. So, it may be better to use the selfadjoint operator and it's spectral decomposition.

Dynamics of distances depends not on the Jacobian, but on the symmetrized Jacobian: $${d(\Delta
\Psi,\Delta \Psi) \over dt}=(\Delta \Psi, [D_{\Psi}J(\Psi)+(D_{\Psi}J(\Psi))^+]\Delta \Psi) +
o(\Delta \Psi),$$ where $(\:,\:)$ is usual scalar product, $\Delta \Psi$ is difference between two
solutions of equation (\ref{sys}), $\Psi=\Psi(t)$ is one of these solutions.

In the theory of inertial manifolds \cite{IneManCFTe88,JonTiti,Chueshov}, for example, one usually
uses the following form of equation (\ref{sys}) with selfadjoint linear operator $A$:
$\dot{\Psi}+A\Psi=R(\Psi),$ and spectral decomposition of $A$ rules the fast-slow splitting.

There are different physically motivated ways to select the scalar product and create the
symmetrization \cite{GKZDPhA2000,InChLANL,CMIM}. But symmetrization does not provide termodinamicity
and the entropy for the projected equations can decrease.

The construction of the thermodynamic projector which always preserve the dissipation is simple and
transparent. We shall describe it now, in the introduction, and it's uniqueness will be proved in the
next section. The proof of uniqueness will demonstrate, that all other ways of projection are
thermodynamically inconsistent, and lead to entropy decrease, and, hence, to the perpetuum mobile of
the second kind.

Let for the system (\ref{sys}) the entropy $S(\Psi)$ exist, and
\begin{equation} \label{entrgrow}
{dS \over dt} = (D_{\Psi}S)J(\Psi) \geq 0 .
\end{equation}
We introduce the {\it entropic scalar product} $\langle \: \mid \: \rangle_{\Psi}$:
\begin{equation}\label{es1}
\langle a \mid b\rangle_{\Psi} = -(a,(D_{\Psi}^2S)(b)),
\end{equation}
where $D_{\Psi}^2S$ is the second differential of the entropy.

The thermodynamic projector is defined for a given point $\Psi$ and a subspace $T$ (the tangent space
to an ansatz manifold). Let us consider a subspace $T_0 \subset T$ which is annulled by the
differential $S$ in the point $\Psi$: $(D_{\Psi}S)T_0=0$.
 If $T_0 = T$, then the
thermodynamic projector is the orthogonal projector on $T$ with respect to the entropic scalar
product $\langle \: |\:\rangle_{\Psi}$. Suppose that $T_0\neq T$. Let $e_g\in T$, $e_g \perp T_0$
with respect to the entropic scalar product $\langle \: \mid \: \rangle_{\Psi}$, and
$(D_{\Psi}S)(e_g)=1$. These conditions define vector $e_g$ uniquely. The projector onto $T$ is
defined by the formula
\begin{equation}\label{ep}
P(J)=P_0(J)+e_g(D_{\Psi}S)(J),
\end{equation}
\noindent where $P_0$ is the orthogonal projector onto
$T_0$ with respect to the entropic scalar product $\langle \: \mid \: \rangle_{\Psi}$.

For example, if $T$ is a finite-dimensional space, then the projector (\ref{ep}) is constructed in
the following way. Let $e_1,..,e_n$ be a basis in $T$, and for definiteness, $(D_{\Psi}S)(e_1)\neq
0$.

\noindent 1) Let us construct a system of vectors

\begin{equation}
b_i=e_{i+1}-\lambda_i e_1, (i=1,..,n-1),
\end{equation}
 \noindent where
$\lambda_i=(D_{\Psi}S)(e_{i+1})/(D_{\Psi}S)(e_{1})$, and hence $(D_{\Psi}S)(b_i)=0$. Thus,
$\{b_i\}_1^{n-1}$ is a basis in $T_0$.

\noindent2) Let us orthogonalize $\{b_i\}_1^{n-1}$ with respect to the entropic scalar product
$\langle \: \mid \: \rangle_{\Psi}$. We get an orthonormal with respect to $\langle \: \mid \:
\rangle_{\Psi}$ basis $\{g_i\}_1^{n-1}$ in $T_0$.

\noindent3) We find $e_g\in T$ from the conditions:
\begin{equation}
\langle e_g \mid g_i \rangle _{\Psi} = 0, (i=1,..,n-1), (D_{\Psi}S)(e_g)=1.
\end{equation}
\noindent and, finally we get
\begin{equation}\label{pfin}
P(J) = \sum_{i=1}^{n-1}g_i \langle g_i\mid J \rangle_{\Psi}+e_g (D_{\Psi}S)(J)
\end{equation}
If $(D_{\Psi}S)(T)=0$, then the projector $P$ is simply the orthogonal projector with respect to the
$\langle \:| \: \rangle_{\Psi}$ scalar product. This is possible if $\Psi$ is the global maximum of
entropy point (equilibrium). Then

\begin{equation}\label{peq}
P(J) = \sum_{i=1}^n{g_i\langle g_i|J \rangle_{\Psi}}, \langle g_i|g_j \rangle_{\Psi} = \delta_{ij}.
\end{equation}

The entropy production for projected vector field (\ref{pfin}) is the same, as for the initial vector
field (\ref{sys}):
\begin{equation}\label{eprod}
(D_{\Psi}S)( P(J)) = (D_{\Psi}S)(e_g)(D_{\Psi}S)(J).
\end{equation}

The significance of the case $(D_{\Psi}S)(T)=0$ may be not clear at the first glance, because such a
state $\Psi$ should be the equilibrium point with $J(\Psi)=0$. Nevertheless, this case is important
as a limit of nonequilibrium $\Psi$, and for discussion of persistence of the Onsager
relations\footnote{The preservation of the Onsager reciprocity relations for projected equations
follows from the requirement of persistence of the {\it sign} of dissipation. This seems surprising,
because these relations do not follow from the entropy grows. It should be stressed, that only the
conditional statement can be proved: if for the initial system hold the Onsager reciprocity
relations, then these relations hold for the projected system.} \cite{Grm} as well, as for the proof
of uniqueness the thermodynamic projector.

In this paper we do not discuss the third main problem of model reduction: How to improve the
constructed manifold and the projector in order to make the manifold more invariant and the motion
along it more slow. This discussion can be found in different works
\cite{GK1,GKTTSP94,Lam,IneManCFTe88,JonTiti,CMIM}.

The discovery of the molecular individualism for  dilute polymers in the flow \cite{Chu} was the
challenge to theory from the very beginning. ``Our data should serve as a guide in developing
improved microscopic theories for polymer dynamics"... was the concluding sentence of the paper
\cite{Chu}. P. de Gennes invented the term ``molecular individualism" \cite{DeGenne}. He stressed
that in this case the usual averaging procedures are not applicable. At the highest strain rates
distinct conformation shapes with different dynamics were observed \cite{Chu}. Further works for
shear flow demonstrated not only shape differences, but different large temporal fluctuations
\cite{Chu2}.

Equation for the molecules in a flow are known. These are the Fokker-Planck equations with external
force. The theory of the molecular individualism is hidden inside these equations. Following the
logic of model reduction we should solve two problems: to construct the slow manifold, and to project
the equation on this manifold. The second problem is solved: the thermodynamic projector is necessary
for this projection. Why should we use this projector also for driven systems? These systems can be
formally written as
\begin{equation} \label{sysdri}
{d\Psi \over dt}= J(\Psi)+J_{\rm ex},
\end{equation}
where  $J_{\rm ex}$ is the external field (driven force).

The entropy for  system (\ref{sysdri}) can decrease, but the thermodynamic processes modeled by the
term $J(\Psi)$ should always produce the entropy (both in the initial and in the projected systems).
This is the reason to use the thermodynamic projector also for open systems.

How to solve the first problem? We can find a hint in the paper \cite{IK00}. The Gaussian
distributions form the invariant manifold for the FENE-P model of polymer dynamics, but, as it was
discovered in \cite{IK00}, this manifold can become unstable in the presence of a flow. We propose to
model this instability as dissociation of the Gaussian peak into two peaks. This dissociation
describes appearance of an unstable direction in the configuration space.

In the classical FENE-P model of polymer dynamics a polymer molecule is represented by one
coordinate: the stretching of molecule (the connector vector between the beads). There exists a
simple mean field generalized models for multidimensional configuration spaces of molecules. In these
models dynamics of distribution functions is described by the Fokker-Planck equation in a quadratic
potential well. The matrix of coefficients of this quadratic potential depends on the matrix of the
second order moments of the distribution function. The Gaussian distributions form the invariant
manifold for these models, and the first dissociation of the Gaussian peak after appearance of the
unstable direction in the configuration space has the same nature and description, as for the
one-dimensional models of molecules considered below.

At the highest strain there can appear new unstable directions, and corresponding dissociations of
Gaussian peaks form a {\it cascade} of dissociation. For $m$ unstable directions we get the Gaussian
parallelepiped: The distribution function is represented as a sum of $2^m$ Gaussian peaks located in
the vertixes of parallelepiped:

\begin{eqnarray}\label{parall}
&&\Psi(q)=\nonumber \\ &&{1 \over 2^m(2\pi)^{n/2}\sqrt{\det \Sigma}} \sum_{\varepsilon_i=\pm 1, \,
(i=1, \ldots, m)} \exp\left(-\frac{1}{2}\left(\Sigma^{-1}\left(q+\sum_{i=1}^m \varepsilon_i
\varsigma_i \right), \: q+\sum_{i=1}^m \varepsilon_i \varsigma_i\right)\right),
\end{eqnarray}
where $n$ is dimension of configuration space, $2\varsigma_i$ is the vector of the $i$th edge of the
parallelepiped, $\Sigma$ is the one peak covariance matrix (in this model $\Sigma$ is the same for
all peaks). The macroscopic variables for this model are:
\begin{enumerate}
\item The covariance matrix $\Sigma$ for one peak;
\item The set of vectors $\varsigma_i$ (or the parallelepiped edges).
\end{enumerate}

The family of distributions (\ref{parall}) can be improved to include the proper equilibrium (this is
important condition: the equilibrium should belong to the ansatz manifold). There may be different
further refinements, some of them are discussed below.

\section{Uniqueness of thermodynamic projector}

In this section, the uniqueness theorem for thermodynamic projector will be proved.

\subsection{Projection of linear vector field}

Let $E$ be a real Hilbert space with the scalar product $\langle \: \mid \: \rangle$, $Q$ be a set of
linear bounded operators in $E$ with negatively definite quadratic form $ \langle Ax \mid x \rangle
\leq 0$ for every $A \in Q$, $T \varsubsetneq E$ be a nontrivial ($T \neq \{0\}$) closed subspace.
For every projector $P:E\rightarrow T$ ($P^2=P$) and linear operator $A:E\rightarrow E$ we define the
projected operator $P(A):T\rightarrow T$ in such a way:
\begin{equation} \label{proop}
P(A) x = PA x \equiv PAP x \: \mbox{for} \: x\in T.
\end{equation}
The space $T$ is the Hilbert space with the scalar product $\langle \: \mid \: \rangle$. Let $Q_T$ be
a set of linear bounded operators in $T$ with negatively define quadratic form $ \langle Ax \mid x
\rangle \leq 0$.

{\bf Proposition 1.} {\it The inclusion $P(Q) \subseteq Q_T$ for a projector $P:E\rightarrow T$ holds
if and only if $P$ is the orthogonal projector with respect to the scalar product $\langle \: \mid \:
\rangle$.}

{\bf Proof.} If $P$ is orthogonal (and, hence, selfadjoint) and $ \langle Ax \mid x \rangle \leq 0$,
then $$ \langle PAPx \mid x \rangle = \langle APx \mid Px \rangle \leq 0.$$

If $P$ is not orthogonal, then $Px \neq 0$  for some vector $x \in T^{\perp}$ in orthogonal
complement of $T$. Let us consider the negatively defined selfadjoint operator $$A_x = -\mid Px-ax
\rangle\langle Px-ax \mid$$ $(A_x y = - (Px-ax)\langle Px-ax \mid y \rangle ).$ The projection of
$A_x$ on $T$ is: $$P(A_x)=(a-1) \mid Px \rangle\langle Px \mid.$$ This operator is not negatively
definite for $a>1$. $\square$

Immediately from this proof follows the Corollary 1.

{\bf Corollary 1.} {\it Let $Q^{\rm sym} \subset Q$ be a subset of selfadjoint operators in $E$. The
inclusion $P(Q^{\rm sym}) \subseteq Q_T$ for a projector $P:E\rightarrow T$ holds if and only if $P$
is the orthogonal projector with respect to the scalar product $\langle \: \mid \: \rangle$.}
$\square$

{\bf Corollary 2.} {\it Let $Q^{\rm sym}_T \subset Q_T$ be a subset of selfadjoint operators in $T$.
If $P(Q) \subseteq Q_T$ for a projector $P:E\rightarrow T$, then $P(Q^{\rm sym}) \subseteq Q^{\rm
sym}_T$.}

It follows from the Proposition 1 and the obvious remark: If operators $A$ and $P$ are selfadjoint,
then operator $PAP$ is selfadjoint too. $\square$

The Proposition 1 means that a projector which transforms every linear vector field $Ax$ with
Lyapunov function $\langle x \mid x \rangle$ into projected vector field $PAPx$ with the same
Lyapunov function is orthogonal with respect to the scalar product $\langle \: \mid \: \rangle$.

According to the Corollary 1, the conditions of the Proposition 1 can be made weaker: A projector
which transforms every {\it selfadjoint} linear vector field $Ax$ with Lyapunov function $\langle x
\mid x \rangle$ into projected vector field $PAPx$ with the same Lyapunov function is orthogonal with
respect to the scalar product $\langle \: \mid \: \rangle$. In physical applications it means, that
we can deal with requirement of dissipation persistence for vector field with Onsager reciprocity
relations. The consequence of such a requirement will be the same, as for the class of all continuous
linear vector field: The projector should be orthogonal.

The Corollary 2 is a statement about persistence of the reciprocity relations.

\subsection{The uniqueness theorem}

In this subsection we will discuss finite-dimensional systems. There are technical details which make
the theory of nonlinear infinite-dimensional case too cumbersome: the Hilbert space equipped with
entropic scalar product  $\langle \: \mid \: \rangle_{\Psi}$ (\ref{eprod}) for different $\Psi$
consists of different functions. Of course, there exists a common dense subspace, and geometrical
sense remains the same, as for the finite-dimensional space, but we prefer to defer the discussion of
all these details till a special mathematical publication.

Let $E$ be $n$-dimensional real vector space, $U \subset E$ be a domain in $E$, and a $m$-dimensional
space of parameters $L$ be defined, $m<n$, and let $W$ be a domain in $L$. We consider differentiable
maps, $F:\ W\rightarrow U$, such that, for every $y\in W$, the differential of $F$, $D_yF:\
L\rightarrow E$, is an isomorphism of $L$ on a subspace of $E$. That is, $F$ are the manifolds,
immersed in the phase space of the dynamic system (\ref{sys}), and parametrized by parameter set $W$.

Let the twice differentiable function $S$ on $U$ be given (the entropy). We assume that $S$ is
strictly concave in the second approximation: The quadratic form defined by second differential of
the entropy $D^2_{\Psi} S (x,x)$ is strictly negative definite in $E$ for every $\Psi \in U$. We will
use the entropic scalar product (\ref{es1}). Let $S$ have the interior point of maximum in $U$:
$\Psi^{eq} \in {\rm int} U.$

The function $S$ is Lyapunov function for a vector field $J$ in $U$, if $(D_{\Psi} S) (J(\Psi)) \geq
0$ for every $\Psi \in U$.

First of all, we shall study vector fields with Lyapunov function $S$ in the neighborhood of
$\Psi^{eq}$. Let $0 \in  {\rm int} W,$ $F:\ W\rightarrow U$ be an immersion, and $F(0) = \Psi^{eq}.$
Let us define $T_y = {\rm im} D_yF(y)$ for each $y \in W.$ This $T_y$ is the tangent space to $F(W)$
in the point $y$. Suppose that the mapping  $F$ is sufficiently  smooth, and $F(W)$ is not tangent to
entropy levels: $$T_y \nsubseteq \ker D_{\Psi}S|_{\Psi=F(y)}$$ for every $y\neq 0$. The thermodynamic
projector for a given $F$ is a projector-valued function $y \mapsto P_y,$ where $P_y: E \rightarrow
T_y$ is a projector. The {\bf thermodynamic conditions} reads: {\it For every smooth vector field
$J(\Psi)$ in $U$ with Lyapunov function $S$ the projected vector field $P_y(J(F(y)))$ on $F(W)$ has
the same Lyapunov function $S(F(y))$.}

Proposition 1 and Corollaries 1, 2 make it possible to prove uniqueness of the thermodynamic
projector for the weakened thermodynamic conditions too: {\it For every smooth vector field $J(\Psi)$
in $U$ with Lyapunov function $S$ and selfadjoint Jacobian operator for every equilibrium point (zero
of $J(\Psi)$) the projected vector field $P_y(J(F(y)))$ on $F(W)$ has the same Lyapunov function
$S(F(y))$.} We shall not discuss it separately.

{\bf Proposition 2.} {\it Let the thermodynamic projector $P_y$ be a smooth function of $y$. Then }
\begin{equation} \label{Prop2}
P_0=P^{\perp}_0 \: \mbox{and} \: P_y=P^{\perp}_y + O(y),
\end{equation}
{\it where $P^{\perp}_y$ is orthogonal projector onto $T_y$ with  respect to the entropic scalar
product $\langle \: |\:\rangle_{F(y)}$.}

{\bf Proof.} A smooth vector field in the neighborhood of $F(0) = \Psi^{eq}$ can be presented as
$A(\Psi - \Psi^{eq}) + o(\|\Psi - \Psi^{eq}\|)$, where $A$ is a linear operator. If $S$ is Lyapunov
function for this vector field, then the quadratic form $ \langle Ax \mid x \rangle_{\Psi^{\rm eq}}$
is negatively definite. $P_y = P_0 + O(y)$, because $P_y$ is a continuous function. Hence, for $P_0$
we have the problem solved by the Proposition 1, and $P_0=P^{\perp}_0$. $\square$

{\bf Theorem 1.} {\it Let the thermodynamic projector $P_y$ be a smooth function of $y$. Then}
\begin{equation} \label{tpro}
P_y=P_{0y}+  e_g D_{\Psi}S|_{\Psi=F(y)},
\end{equation}
{\it where notations of formula (\ref{ep}) are used: $T_{0y}$ is the kernel  of linear functional
$D_{\Psi}S|_{\Psi=F(y)}$ in $T_y$, $P_{0y}: T_{0y} \rightarrow E$ is the orthogonal projector with
respect to the entropic scalar product  $\langle \: \mid \: \rangle_{F(y)}$ (\ref{eprod}). Vector
$e_g \in T $ is proportional to the Riesz representation $g_y$ of linear functional
$D_{\Psi}S|_{\Psi=F(y)}$ in $T_y$ with respect to the entropic scalar product: $$\langle g_y \mid x
\rangle_{F(y)} = (D_{\Psi}S|_{\Psi=F(y)})(x)$$ for every $x \in T_y $, $e_g = g_y / \langle g_y \mid
g_y \rangle_{F(y)}$.}

{\bf Proof.} Let $y\neq 0$. Let us consider auxiliary class of vector fields $J$ on $U$ with
additional linear balance $(D_{\Psi}S)_{\Psi=F(y)})(J)=0$. If such a vector field has Lyapunov
function $S$, then $\Psi=F(y)$ is its equilibrium point: $J(F(y))=0$. The class of vector fields with
this additional linear balance and Lyapunov function $S$ is sufficiently rich and we can use the
Propositions 1, 2 for dynamics on the auxiliary phase space $$\{x \in U |
(D_{\Psi}S|_{\Psi=F(y)})(x-F(y))=0\}.$$ Hence, the restriction of $P_y$ on the hyperplane $\ker
D_{\Psi}S|_{\Psi=F(y)}$ is $P_{0y}$. Formula (\ref{tpro}) gives the unique continuation of this
projector on the whole $E$. $\square$

\subsection{Orthogonality of the thermodynamic projector and entropic gradient models}

In Euclidean spaces with  given scalar product, we often identify the differential of a function
$f(x)$ with its gradient: in orthogonal coordinate system   $({\rm grad} f(x))_i = \partial f(x) /
\partial x_i $. However, when dealing with a more general setting, one can run into problems making
sense out of such a definition. What to do, if there is no distinguished scalar product, no given
orthogonality?

For a given scalar product $\langle \: | \: \rangle$ the gradient ${\rm grad}_x f(x)$ of a function
$f(x)$ at a point $x$ is such a vector $g$ that $\langle g | y\rangle = D_xf (y)$ for any vector $y$,
where $D_xf$ is the differential of function $f$ at a point $x$. The differential of function $f$ is
the linear functional that provides the best linear approximation near the given point.

To transform a vector into a linear functional one needs a {\it pairing}, that means a bilinear form
$\langle \: | \: \rangle$. This pairing transforms vector $g$ into linear functional $\langle g|$:
 $\langle g|(x)=\langle g|x\rangle$. Any twice differentiable function $f(x)$ generates a field of pairings:
at any point $x$ there exists a second differential of $f$, a quadratic form $(D^2_x f)(\Delta
x,\Delta x)$. For a convex function these forms are positively defined, and we return to the concept
of scalar product. Let us calculate a gradient of $f$ using this scalar product. In coordinate
representation
\begin{equation}\label{New}
\sum_{i}({\rm grad} f(x))_i \frac {\partial^2 f}{\partial x_i \partial x_j}= \frac {\partial
f}{\partial x_j}, \: \mbox{hence,} \: ({\rm grad} f(x))_i = \sum_{i} (D^2_x f)^{-1}_{ij}\frac
{\partial f}{\partial x_j}.
\end{equation}
As we can see, this ${\rm grad} f(x)$ is the $Newtonian$ direction, and with this gradient the method
of steepest descent transforms into the Newton's method of optimization.

Entropy is the concave function and we defined the entropic scalar product through negative second
differential of entropy (\ref{es1}). Let us define the gradient of entropy by means of this scalar
product: $\langle{\rm grad}_{\Psi} S|x\rangle_{\Psi} = (D_{\Psi}S)(x)$. The {\it entropic gradient
system} is
\begin{equation}\label{entgrad}
\frac{d\Psi}{dt}=\varphi({\Psi}){\rm grad}_{\Psi} S,
\end{equation}
where $\varphi({\Psi})>0$ is a positive kinetic multiplier.

The system (\ref{entgrad}) is a representative of a family of {\it model kinetic equations}. One
replaces complicated kinetic equations by model equations for simplicity. The main requirements to
such models are: they should be as simple as possible and should not violate the basic physical laws.
The most known model equation is the BGK model \cite{BGK} for substitution of collision integral in
the Boltzmann equation. There are different models for simplifying kinetics \cite{GKMod,LFHMod}. The
entropic gradient models (\ref{entgrad}) possesses all the required properties (if the entropy
Hessian is sufficiently simple). It was invented first for Lattice-Boltzmann kinetics \cite{AKMod}.
In many cases it is more simple than the BGK model, because the gradient model is {\it local} in the
sense that it uses only the entropy function and its derivatives at a current state, and it is not
necessary to compute the equilibrium (or quasiequilibrium for quasiequilibrium models
\cite{InChLANL,GKMod}. The entropic gradient model has an one-point relaxation spectrum, because the
gradient vector field (\ref{entgrad}) has near an equilibrium $\Psi^{eq}$ an extremely simple linear
approximation: $d(\Delta \Psi)/dt=-\varphi({\Psi^{eq}})\Delta \Psi$. It corresponds to a well-known
fact that the Newton's method minimizes a positively defined quadratic form in one step.

Direct calculation shows that the thermodynamic projector $P$ (\ref{ep}) in a point $\Psi$ onto the
tangent space $T$ can be rewritten as
\begin{equation}\label{projgrad}
P(J) = P^{\bot} (J) + \frac{ {\rm grad}_{\Psi} S^{\|}}{\langle {\rm grad}_{\Psi} S^{\|}|{\rm
grad}_{\Psi} S^{\|}\rangle_{\Psi}} \langle {\rm grad}_{\Psi} S^{\bot}|J \rangle_{\Psi},
\end{equation}
where $P^{\bot}$ is the orthogonal projector onto $T$ with respect the entropic scalar product, and
the gradient ${\rm grad}_{\Psi} S$ is splitted onto tangent and orthogonal components: $${\rm
grad}_{\Psi} S = {\rm grad}_{\Psi} S^{\|} + {\rm grad}_{\Psi} S^{\bot}; \: {\rm grad}_{\Psi} S^{\|}=
P^{\bot} {\rm grad}_{\Psi} S;  \: {\rm grad}_{\Psi} S^{\bot} = (1-P^{\bot}) {\rm grad}_{\Psi} S .$$

From Eq. (\ref{projgrad}) it follows that two properties of an ansatz manifolds are equivalent:
orthogonality of the thermodynamic projector and invariance of the manifold with respect to the
entropic gradient system (\ref{entgrad}).

{\bf Proposition 3.} {\it The thermodynamic projector for an ansatz manifold $\Omega$  is orthogonal
at any point ${\Psi} \in \Omega$ if and only if ${\rm grad}_{\Psi} S \in T_{\Psi}(\Omega)$ at any
point ${\Psi} \in \Omega$.} $\square$

It should be stressed that it should be possible to think of gradients as infinitesimal displacements
of points $\Psi$.  Usually there are some balances, at least the conservation of the total
probability, and the gradient should belong to a given subspace of zero balances change. For example,
for the classical Boltzmann-Gibbs-Shannon entropy $S=- \int \Psi(q) (\ln \Psi(q) - 1)d^n q$ the
entropic scalar product is $\langle g(q)|f(q)\rangle_{\Psi}= \int g(q)f(q)/\Psi(q) d^n q$ and ${\rm
grad}_{\Psi} S = - {\Psi(q)}\ln ({\Psi(q)}) + c(q)$, where function (vector) $c(q)$ is orthogonal to
a given subspace of zero balances. This function have to be founded from the conditions of zero
balances for the gradient ${\rm grad}_{\Psi} S$. For example, if the only balance is the conservation
of the total probability, $\int \Psi(q) d^n q \equiv 1 $, then for the classical
Boltzmann-Gibbs-Shannon entropy $S$
\begin{equation}
{\rm grad}_{\Psi} S = - \Psi(q)\left(\ln(\Psi(q)) - \int \Psi(q')\ln(\Psi(q')) d^n q'\right).
\end{equation}
For the Kullback-form entropy (i.e. for the negative free energy or the Massieu-Planck functions)
$$S=-F/T = - \int \Psi(q) \left(\ln \left(\frac{\Psi(q)}{\Psi^{eq}(q)}\right) - 1 \right)d^n q$$ the
second differential and the entropic scalar product are the same, as for the classical
Boltzmann-Gibbs-Shannon entropy, and
\begin{equation}
{\rm grad}_{\Psi} S = - \Psi(q)\left(\ln \left(\frac{\Psi(q)}{\Psi^{eq}(q)}\right) - \int \Psi(q')
\ln \left(\frac{\Psi(q)}{\Psi^{eq}(q)}\right)d^n q'\right).
\end{equation}
For more complicated system of balances, linear or non-linear, the system of linear  equations for
$c(q)$ can also be written explicitly.

\subsection{Violation of transversality condition, singularity of thermodynamic projection and steps
of relaxation}

The thermodynamic projector transforms the arbitrary vector field equipped with the given Lyapunov
function into a vector field with the same Lyapunov function for a given ansatz manifold which is not
tangent to the Lyapunov function levels. Sometimes it is useful to create an ansatz with violation of
this transversality condition. The point of entropy maximum on this ansatz is not the equilibrium.
The usual examples are: the non-correlated approximation $\Psi(x_1,\ldots, x_n)=\prod_i f(x_i)$, the
Gaussian manifold, etc. For these manifolds the thermodynamic projector  becomes singular near the
point of entropy maximum $\Psi^*$ on the ansatz manifold. It is obvious from Eq. (\ref{projgrad}): in
the neighborhood of $\Psi^*$ it has the form
\begin{eqnarray}\label{sing}
P(J) &=& P^{\bot} (J) + \frac{ {\rm grad}_{\Psi} S^{\|}}{\langle {\rm grad}_{\Psi} S^{\|}|{\rm
grad}_{\Psi} S^{\|}\rangle_{\Psi}} \langle {\rm grad}_{\Psi} S^{\bot}|J \rangle_{\Psi} = \nonumber \\
&& -\frac{ \Delta \Psi}{\langle\Delta \Psi| \Delta \Psi\rangle_{\Psi^*}} \sigma (\Psi^*) + O(1),
\end{eqnarray}
where $\Delta \Psi = \Psi - \Psi^*$ is the deviation of $\Psi$ from  $\Psi^*$, $\sigma
(\Psi^*)=\langle {\rm grad}_{\Psi^*} S^{\bot}|J \rangle_{\Psi^*}$ is the entropy production at the
point $\Psi^*$, $\sigma (\Psi^*) \neq 0 $, because the point of entropy maximum  $\Psi^*$ is not the
equilibrium. In this case the projected system in the neighborhood of $\Psi^*$ reaches the point
$\Psi^*$ at finite time $t^*$ as $ \sqrt{t^*-t}$. The entropy difference $\Delta S=S(\Psi)-S(\Psi^*)=
-\frac{1}{2}{\langle\Delta \Psi| \Delta \Psi\rangle_{\Psi^*}} + o({\langle\Delta \Psi| \Delta
\Psi\rangle_{\Psi^*}}) $ goes to zero as $-\sigma (\Psi^*)(t^*-t)$ ($t \leq t^*$).

The singularity of projection has a transparent physical sense. The relaxation along the ansatz
manifold to the point $\Psi^*$ is not complete, because this point is not the equilibrium. This
motion should be rated as a step of relaxation, and after it was completed, the next step should
start. In this sense it is obvious that the motion to the point $\Psi^*$ along the ansatz manifold
should take the finite time. The results of this step-by-step relaxation can represent the whole
process (with smoothing \cite{Z1}, or without it \cite{Z2}). The experience  of such step-by-step
computing of relaxation trajectories in the initial layer problem for the Boltzmann kinetics
demonstrated its efficiency \cite{Z1,Z2}.

\subsection{Thermodynamic projector,  quasiequilibrium and entropy maximum}

The thermodynamic projector projects any vector field which satisfies the second law of
thermodynamics into the vector field which satisfies the second law too. Another projectors violate
the second law. But what does it mean? Each projector $P_{\Psi}$ onto tangent space to an ansatz
manifold in a point $\Psi$ induces the fast-slow motion splitting: Fast motion is the motion parallel
to $\ker P_{\Psi}$ (on the affine subspace $\Psi+\ker P_{\Psi}$ in the neighborhood of $\Psi$), slow
motion is the motion on the slow manifold and in the first order it is parallel to the tangent space
$T_{\Psi}$ in the point $\Psi$ (in the first order this slow manifold is the affine subspace
$\Psi+{\rm im} P_{\Psi}$, $T_{\Psi}={\rm im} P_{\Psi}$), and velocity of the slow motion in point
$\Psi$ belongs to image $P_{\Psi}$.

If $P_{\Psi}$ is the thermodynamic projector, then $\Psi$ is the point of entropy maximum on the
affine subspace of fast motion $\Psi+\ker P_{\Psi}$. It gives the solution to the problem
\begin{eqnarray}\label{maxfast}
S(x)\rightarrow \max, \:  x\in \Psi+\ker P_{\Psi}.
\end{eqnarray}
This is the most important property of thermodynamic projector. It was introduced in our paper
\cite{GK1} as a main thermodynamic condition for model reduction. Let us call it for nonequilibrium
points $\Psi$ {\it the property } ${\bf A}$:
\begin{eqnarray}\label{condA}
{\bf A.} &\:& \ker P_{\Psi} \subset \ker D_{\Psi} S.
\end{eqnarray}

If the projector $P_{\Psi}$ with the property ${\bf A}$ can be continued to the equilibrium point,
$\Psi^{\rm eq}$, as a smooth function of $\Psi$, then in this point $\ker P_{\Psi} \perp {\rm im}
P_{\Psi}$. If this is valid for all systems (including systems with additional linear balances), then
the following {\it property} ${\bf B}$ holds:
\begin{eqnarray}\label{condB}
{\bf B.} &\:& (\ker P_{\Psi} \bigcap \ker D_{\Psi} S)  \perp ({\rm im} P_{\Psi} \bigcap \ker D_{\Psi}
S).
\end{eqnarray}
Of course, orthogonality in formulae (\ref{condA},\ref{condB}) is considered  with respect to the
entropic scalar product in point $\Psi$.

The property ${\bf A}$ means that the value of entropy production persists for all nonequilibrium
points. The sense of property ${\bf B}$ is: each point of the slow manifold can be made an
equilibrium point (after the deformation of the system which leads to appearance on additional
balance). And for equilibrium points the orthogonality condition (\ref{condB}) follows from the
property ${\bf A}.$

If $P_{\Psi}$ does not have the property ${\bf A}$, then $\Psi$ is not the point of entropy maximum
on the affine subspace of fast motion $\Psi+\ker P_{\Psi}$, so either the fast motion along this
subspace does not leads to $\Psi$ (and, hence, the point $\Psi$ does not belong to slow manifold), or
this motion violates the second law, and the entropy decreases. This is the violation of the second
law of thermodynamics during the fast motion. If $P_{\Psi}$ does not have  the property ${\bf A}$,
then such a violation is expected for almost every system.

On the other hand, if $P_{\Psi}$ is not the thermodynamic projector, then there exists a
thermodynamic vector field $J$, with non-thermodynamic projection: $S$ is Lyapunov function for $J$
(it increases), and is not Lyapunov function for $P_{\Psi}(J)$ (it decreases in the neighborhood of
$\Psi$). The difference between violation of the second law of thermodynamics in fast and slow
motions for a projector without the property ${\bf A}$ is: for the fast motion this violation
typically exists, for the slow (projected) motion there exist some thermodynamic systems with such a
violation. On the other hand, the violation in slow motion is more important for applications, if we
use the slow dynamics as an answer (and assume that the fast dynamics is relaxed).

If $P_{\Psi}$ does not have the property ${\bf B}$, then there exist systems with violation of the
second law of thermodynamics in fast and slow motions. Here we can not claim that the second law
violates for almost every system, but such systems exist.

One particular case of thermodynamic projector is known during several decades. It is the
quasiequilibrium projector on the tangent space of the quasiequilibrium (MaxEnt) manifold.

Let a set of macroscopic (slow) variables be given: $M=m(\Psi)$. The vector of macroscopic variables
$M$ is a continuous linear function of microscopic variables $\Psi$. Let the ansatz manifold be the
manifold of conditional entropy maximum:
\begin{eqnarray}\label{maxent}
S(\Psi)\rightarrow \max, \:  m(\Psi)=M.
\end{eqnarray}
The solution of the problem (\ref{maxent}) $\Psi^{\rm qe}_M$ parametrized by values of the
macroscopic variables $M$ is the quasiequilibrium manifold.

The projector on the tangent space to the quasiequilibrium manifold is:
\begin{eqnarray}\label{qepro}
\pi^{\rm qe}_M=\left(D_M\Psi^{\rm qe}_M\right)_M m=   \left(D_{\Psi}^2S \right)_{\Psi^{\rm
qe}_M}^{-1}m^T \left(m\left(D_{\Psi}^2S\right)_{\Psi^{\rm qe}_M}^{-1}m^T\right)^{-1}m.
\end{eqnarray}
This formula was essentially obtained by Robertson \cite{Robertson}\footnote{In his dissertation
\cite{Robertson} B. Robertson has studied ``the equation of motion for the generalized canonical
density operator". The generalized canonical density renders entropy a maximum for given statistical
expectations of the thermodynamic coordinates. He started from the Liouville equation for a general
quantum system. The first main result of Robertson's paper  is the explicit expression for splitting
of the motion onto two components: projection of the motion onto  generalized canonical density and
the motion in the kernel of this projection. The obtained projector operator is a specific particular
case of the quasiequilibrium projector (\ref{qepro}). The second result is exclusion of the motion in
the kernel of quasiequilibrium projector from the dynamic equation. This operation is similar to the
Zwanzig formalism \cite{Zwa}. It leads to the integro-differential equation with delay in time for
the generalized canonical density. The quasiequilibrium projector (\ref{qepro}) is more general than
the projector obtained by Robertson \cite{Robertson} in the following sense: It is derived for any
functional $S$ with non-degenerate second differential $D_{\Psi}^2S$, for manifold of condition
maxima of $S$ and for any (nonlinear) differential equation. B. Robertson emphasized that this
operator is non-Hermitian with respect to standard $L^2$ scalar product and in this sense is not a
projector at all. Nevertheless it is  self-adjoint (and, hence, orthogonal), but with respect to
another (entropic) scalar product. The general thermodynamic projector (\ref{ep}) acts with an
arbitrary ansatz manifolds and in that sense is much more general. The motion in transversal
direction will be discussed below together with post-processing algorithms.}.

First of all, the thermodynamic projector (\ref{qepro}) for the quasiequilibrium manifold
(\ref{maxent}) is the orthogonal projector with respect to the entropic scalar product (\ref{es1}).
In this case both terms in the thermodynamic projector (\ref{ep}) are orthogonal projectors with
respect to the entropic scalar product (\ref{es1}). The first term, $P_0$, is orthogonal projector by
construction. For the second term, $e_g (D_{\Psi}S)$, it means that the Riesz representation of the
linear functional $D_{\Psi}S$ in the whole space $E$ with respect to the entropic scalar product
belongs to the tangent space of the quasiequilibrium manifold. This Riesz representation is the
gradient of $S$ with respect to $\langle \: | \: \rangle_{\Psi}$. The following Proposition gives
simple and important condition of orthogonality of the thermodynamic projector (\ref{ep}). Let
$\Omega$ be an ansatz manifold, and let $\VV$ be some quasiequilibrium manifold, $\Psi \in \Omega
\bigcap \VV$, $T_{\Psi}$ be the tangent space to the ansatz manifold $\Omega$ in the point $\Psi$.
Suppose that there exists a neighborhood of $\Psi$ where $\VV \subseteq \Omega$. We use the notation
${\rm grad}_{\Psi}S$ for the Riesz representation of the linear functional $D_{\Psi}S$ in the
entropic scalar product $\langle \: | \: \rangle_{\Psi}$: $\langle{\rm grad}_{\Psi}S | f
\rangle_{\Psi} \equiv (D_{\Psi}S)(f)$ for $f \in E$.

{\bf Proposition 4.} {\it Under given assumptions, ${\rm grad}_{\Psi}S \in T_{\Psi}$, and the
thermodynamic projector $P_{\Psi}$ is the orthogonal projector onto $T_{\Psi}$ with respect to the
entropic scalar product (\ref{es1}).} $\square$

So, if a point $\Psi$ on the ansatz manifold $\Omega$ belongs to some quasiequilibrium  submanifold
$\VV \subseteq \Omega$, then the thermodynamic projector in this point is simply the orthogonal
projector with respect to the entropic scalar product (\ref{es1}).

Proposition 4 is useful in the following situation. Let the quasiequilibrium approximation be more or
less satisfactory, but the ``relevant degrees of freedom" depend on the current state of the system.
It means that for some changes of the state we should change the list of relevant macroscopic
variables (moments of distribution function for generation the quasiequilibrium). Sometimes it can be
described as presence of hidden degrees of freedom, which are not moments. In these cases the
manifold of reduced description should be extended. We have a family of systems of moments
$M_{\alpha}=m_{\alpha}(\Psi)$, and a family of corresponding quasiequilibrium manifolds
$\Omega_{\alpha}$: The manifold $\Omega_{\alpha}$ consist of solutions of optimization problem
$S(\Psi) \rightarrow \max$, $m_{\alpha}(\Psi)=M$ for given  $\alpha$ and all admissible values for
$M$. To create a manifold of reduced description it is possible to join all the moments $M_{\alpha}$
in one family, and construct the corresponding quasiequilibrium manifold. Points on this manifold are
parametrized by the family of moments values $\{M_{\alpha}\}$ for all possible $\alpha$. It leads to
a huge increase of the quasiequilibrium manifold. Another way to extension of the quasiequlibrium
manifold is a union of all the manifolds $\Omega_{\alpha}$ for all $\alpha$. In accordance with the
Proposition 4,  the thermodynamic projector for this union is simply the orthogonal projector with
respect to the entropic scalar product. This kind of manifolds gives a closest  generalization of the
quasiequilibrium manifolds. An example of such a construction will be described below.

Quasiequilibrium approximation became very popular after works of Jaynes \cite{Janes1}\footnote{From
time to time it is  discussed in the literature, who was the first to introduce the quasiequilibrium
approximations, and how to interpret them. At least a part of the discussion is due to a different
role the quasiequilibrium plays in the entropy--conserving and the dissipative dynamics. The very
first use of the entropy maximization dates back to the classical work of G.\ W.\ Gibbs \cite{Gibb},
but it was first claimed for a principle by E.\ T.\ Jaynes  \cite{Janes1}. Probably the first
explicit and systematic use of quasiequilibria to derive dissipation from entropy--conserving systems
is due to the works of D.\ N.\ Zubarev. Recent detailed exposition is given in \cite{Zubarev}. For
dissipative systems, the use of the quasiequilibrium to reduce description can be traced to the works
of H.\ Grad on the Boltzmann equation \cite{Grad}. The viewpoint of the present authors was
influenced by the papers by L.\ I.\ Rozonoer and co-workers, in particular, \cite{KoRoz,Ko,Roz}. A
detailed exposition of the quasiequilibrium approximation for Markov chains is given in the book
\cite{G1} (Chapter 3, {\it Quasiequilibrium and entropy maximum}, pp.\ 92-122), and for the BBGKY
hierarchy in the paper \cite{Kark}. We have applied maximum entropy principle to the description the
universal dependence the 3-particle distribution function $F_3$ on the 2-particle distribution
function $F_2$ in classical systems with binary interactions \cite{BGKTMF}. A general discussion of
the maximum entropy principle with applications to dissipative kinetics is given in the review
\cite{Bal}. The methods for corrections the quasiequilibrium approximations are developed in
\cite{GK1,GKTTSP94,KTGOePhA2003,Plenka}}. Due to Eq. (\ref{maxfast})  the thermodynamic projector
gives the presentation of almost arbitrary ansatz as the quasiequilibrium manifold. This property
opens the natural field for applications of thermodynamic projector: construction of Galerkin
approximations with thermodynamic properties.

Of course, there is a ``law of the difficulty conservation": for the quasiequilibrium with the moment
parameterization the slow manifold is usually not  explicitly given, and it can be difficult to
calculate it. Thermodynamic projector completely eliminates this difficulty: we can use almost any
manifold as appropriate ansatz now. On the other side, on the quasiequilibrium manifold with the
moment parameterization (if it is found) it is easy to find the dynamics: simply write
$\dot{M}=m(J)$. The building of the thermodynamic projector may require some efforts. Finally, if the
classical quasiequilibrium manifold is found, then it is easy to find the projection of any
distributions $\Psi$ on the quasiequilibrium manifold: $\Psi \mapsto m(\Psi) \mapsto \Psi^{\rm
qe}_{m(\Psi)}$. It requires just a calculation of the moments $m(\Psi)$. The preimage of the point
$\Psi^{\rm qe}_{m(\Psi)}$ is a set (an affine manifold) of distributions $\{\Psi | m(\Psi - \Psi^{\rm
qe}_{m(\Psi)})=0 \}$, and $\Psi^{\rm qe}_{m(\Psi)}$ is the point of entropy maximum on this set. It
is possible, but not so easy, to construct such a projector of some neighborhood of the manifold
$\Omega$ onto $\Omega$ for general thermodynamic projector $P_{\Psi}$ too: for a point $\Phi$ from
this neighborhood
\begin{equation}\label{glopro}
\Phi \mapsto \Psi \in \Omega, \: \: \mbox{if} \: \: P_{\Psi}(\Phi-\Psi)=0.
\end{equation}
A point $\Psi \in \Omega$ is the point of entropy maximum on the preimage of $\Psi$, i.e. on the
affine manifold $\{\Phi | P_{\Psi}(\Phi-\Psi)=0 \}$. It is necessary to emphasize that the map
(\ref{glopro}) can be defined only in a neighborhood of the manifold $\Omega$, but not in the whole
space, because some of affine subspaces $\{\Phi | P_{\Psi}(\Phi-\Psi)=0 \}$ for different $\Psi \in
\Omega$ can intersect. Let us introduce a special denotation for projection of some neighborhood of
the manifold $\Omega$ onto $\Omega$, associated with the thermodynamic projector $P_{\Psi}$
(\ref{glopro}): ${\bf P}_{\Omega}: \Phi \mapsto \Psi$. The preimage of a point $\Psi \in \Omega$ is:
\begin{equation}\label{ProPre}
{\bf P}_{\Omega}^{-1} \Psi = \Psi+\ker P_{\Psi},
\end{equation}
(or, strictly speaking, a vicinity of $\Psi$ in this affine manifold). Differential of the operator
${\bf P}_{\Omega}$ at a point $\Psi \in \Omega$ from the manifold $\Omega$ is simply the projector
$P_{\Psi}$:
\begin{equation}\label{PorDif}
{\bf P}_{\Omega}(\Psi+\varepsilon \Phi)= \Psi + \varepsilon P_{\Psi} \Phi + o(\varepsilon).
\end{equation}
Generally, differential ${\bf P}_{\Omega}$ at a point $\Psi$ has not so simple form, if $\Psi$ does
not belong $\Omega$.

 The ``global extension" ${\bf P}_{\Omega}$ of a field of ``infinitesimal" projectors
$P_{f}$ $(f \in \Omega)$ is needed for discussion of projector operators technique, memory functions
and short memory approximation below.

Is it necessary to use the thermodynamic projector everywhere? The persistence of dissipation is
necessary, because the violation of the second law may lead to strange non-physical effects. If one
creates a very accurate method for solution of initial equation (\ref{sys}), then it may be possible
to expect that the persistence of dissipation will hold without additional efforts. But this
situation yet have not appeared. All methods of model reduction need a special tool to control the
persistence of dissipation.

In order to summarize, let us give three reasons to use the thermodynamic projector:
\begin{enumerate}
\item{It guarantees the persistence of dissipation: all the thermodynamic processes
which should product the entropy conserve this property after projecting, moreover, not only the sign
of dissipation conserves, but the value of entropy production and the reciprocity relations too;}
\item{The coefficients (and, more generally speaking, the right hand part) of kinetic equations are
known significantly worse then the thermodynamic functionals, so, the {\it universality} of the
thermodynamic projector (it depends only on thermodynamic data) makes the thermodynamic properties of
projected system as reliable, as for the initial system;}
\item{It is easy (much more easy than spectral projector, for example).}
\end{enumerate}

\section{Post-processing, memory and natural projector}

\subsection{How to evaluate the ansatz?}

Thermodynamic projector transforms almost arbitrary ansatz into thermodynamically consistent model.
So, the simplest criteria of quality of an ansatz (entropy grows, reciprocity relations, etc.) are
satisfied by the construction of the projector. How to evaluate the ansatz now?

First of all, we can estimate the {\it defect of invariance} $\Delta = J(\Psi)-P_{\Psi}(J(\Psi)).$ If
$\Delta$ is not small (in comparison with the typical value of $J$), then the ansatz should be
improved (for details see, for example, \cite{GKIOeNONNEWT2001,KIOeDev}). It is possible to use
$\Delta$ for error estimation and correction of an ansatz {\it after} solution of projected equations
too (it is so-called post-processing \cite{GaAr,CMIM}). Let $\Psi^0(t), \: (t \in [0,T])$ be the
solution of projected equations $ d\Psi(t) / dt = P_{\Psi}(J(\Psi)),$ and $\Delta(t) =
J(\Psi^0(t))-P_{\Psi^0(t)}(J(\Psi^0(t))).$ Then the following formula
\begin{equation}\label{Picard}
\Psi^1(t)=\Psi^0(t)+ \int_0^t \Delta(\tau) d \tau
\end{equation}
gives the Picard iteration for solution of the initial kinetic equation $ d\Psi(t) / dt = J(\Psi),$
with initial approximation $\Psi^0(t).$ The integral in the right hand side of equation
(\ref{Picard}) gives the estimation of the deviation the ansatz solution $\Psi^0(t)$ from the true
solution as well, as the correction for this ansatz solution. For a better estimation we can take
into account not only  $\Delta(t)$, but the linear part of the vector field $J(\Psi)$ near
$\Psi^0(t)$, and use different approximations of this linear part \cite{CMIM}. The following
representation gives us one of the simplest approximations: $\Psi^1(t)=\Psi^0(t)+\delta \Psi(t);$
\begin{equation}\label{hybrid}
\frac{d(\delta
\Psi(t))}{dt}=\Delta(t)+\frac{\langle\Delta(t)|(DJ)_{\Psi^0(t)}\Delta(t)\rangle_{\Psi^0(t)}}
{\langle\Delta(t)|\Delta(t)\rangle_{\Psi^0(t)}}\delta \Psi(t).
\end{equation}
where $\Delta(t)=J(\Psi^0(t))-P_{\Psi^0(t)}(J(\Psi^0(t))),$ $(DJ)_{\Psi^0(t)}$ is the differential of
$J(\Psi(t))$ in the point $\Psi^0(t)$, $\langle \: |  \: \rangle_{\Psi^0(t)}$ is the entropic scalar
product (\ref{es1}) in the point $\Psi^0(t)$.

The solution of equation (\ref{hybrid}) is
\begin{equation}\label{estim}
 \delta \Psi(t)=
\int_0^t  \exp\left(\int_{\tau}^t k(\theta)d\theta \right)\Delta(\tau)  d \tau,
\end{equation}
where $$k(t)=\frac{\langle\Delta(t)|(DJ)_{\Psi^0(t)}\Delta(t)\rangle_{\Psi^0(t)}}
{\langle\Delta(t)|\Delta(t)\rangle_{\Psi^0(t)}}.$$ The right hand side of equation  (\ref{estim})
improves the simplest Picard iteration (\ref{Picard}) and gives both the estimation of the error of
the ansatz, and the correction for the solution $\Psi^0(t)$.

The projection of $\Delta$ on the slow motion ansatz is zero, hence, for post-processing analysis of
the slow motion, the estimation (\ref{estim}) should be supplemented by one more Picard iteration:
\begin{equation}\label{slowestim}
\delta \Psi_{sl}(t)= P_{\Psi^0(t)} \delta \Psi(t) + \int_0^t
P_{\Psi^0(\tau)}((DJ)_{\Psi^0(\tau)})\delta \Psi(\tau)d\tau,
\end{equation}
where $\delta \Psi(t)$ is calculated by formula (\ref{estim}).

\subsection{Short memory and natural projector}

At the middle of XX century S. Nakajima (1958), R. Zwanzig (1960), and H. Mori (1965) discovered a
new approach to model reduction in statistical physics: the method of projection operators. (Relevant
bibliography and detailed  presentation of this technic can be found in two books
\cite{Eva,Grabert}.) In this section we contrast our approach with the Nakajima--Zwanzig--Mori theory
of projection operators. This theory is based upon two technical steps: (i) a projection technique
for creation exact integro-differential equations that describe dynamics of ``relevant" variables for
given initial conditions, (ii) and various Marcovian, short memory, adiabatic, or other assumptions
of this type that allow us to simplify the exact integro-differential equations. Without such a
simplification the theory is simply equivalent to the initial detailed microscopic dynamics.

The projection operators approach (it is more adequate to call it ``the memory function approach"
\cite{Eva}) deals with linear equations of microscopic description: Liouville equation (or
generalized quantum Liouville equation). The thermodynamic projector developed in this paper and a
series of previous works \cite{GK1,GKTTSP94,InChLANL,Z1} can be applied to any system with entropy,
i.e. to a system with a specified function (functional) whose time derivative should be preserved in
model reduction. Of course, it can be applied to the systems considered in the memory function
approach to project the initial equations onto the manifold of relevant distribution functions.
Moreover, the particular case of the thermodynamic projector, namely the quasiequilibrium projector,
was developed for this purpose by B. Robertson in the context of Nakajima -- Zwanzig approach
\cite{Robertson}.

In this paper we have discussed only the simple infinitesimal projection of the vector field onto the
tangent space to ansatz manifold, but it is possible to develop a hierarchy of  short-memory
approximations even for general nonlinear equations and an arbitrary ansatz manifold
\cite{GKIOeNONNEWT2001,GKOeTPRE2001,GKMex2001,GKPRE02,GKGeoNeo,Plenka,KTGOePhA2003}. This approach
joins ideas of Ehrenfests on coarse-graining \cite{Ehrenfest}, methods of projection operators, and
methods of invariant manifold \cite{GK1,GKTTSP94,InChLANL}. The essence of this approach can be
formulated very simply:  we turn from infinitesimal projection of vector fields to ``natural
projection" of segments of trajectories.

Let us consider a dynamical system (\ref{sys}) $d\Psi / dt= J(\Psi)$, an ansatz manifold $\Omega$, a
field of projectors on $\Omega$: $P_{\Psi}:E \rightarrow T_{\Psi}$ for $\Psi \in \Omega$, and a
global extension ${\bf P}_{\Omega}$ (\ref{glopro}), (\ref{ProPre}) of this field of infinitesimal
projectors. In this construction the field of projectors $P_{\Psi}$ is arbitrary, and thermodynamic
condition will be necessary only for estimation of the entropy production (see the next subsection).
Let ${\bf T}_t$ be the shift in time $t$ due to the dynamical system (\ref{sys}) (the phase flow). We
are looking for a phase flow ${\bf \Theta}_t$ on $\Omega$ that should be a coarse-grained dynamical
system in a short memory approximation. The matching equation for the short memory approximation is
\begin{equation}\label{ShMem}
{\bf \Theta}_{\tau}(\Psi) = {\bf P}_{\Omega}({\bf T}_{\tau}(\Psi)) \: \: \mbox{for all} \: \: \Psi
\in \Omega,
\end{equation}
where $\tau>0$ is the time of memory (it may be a function of $\Psi$: $\tau=\tau (\Psi)$).

The phase flow ${\bf \Theta}_t$ on $\Omega$ is generated by a vector field on $\Omega$:
\begin{equation}\label{VecFi}
{d\Psi \over dt} = \Upsilon (\Psi) = \left.{d {\bf \Theta}_{\tau}(\Psi) \over d\tau }\right|
_{\tau=0} \in T_{\Psi}.
\end{equation}

For the short memory it may be natural to seek a vector field $\Upsilon (\Psi)$ on the ansatz
manifold $\Omega$ in a form of Taylor series in powers of $\tau$: $\Upsilon (\Psi)= \Upsilon_0 (\Psi)
+ \tau \Upsilon_1 (\Psi) +\ldots \: $. Let us expand the right hand side of the matching equation
(\ref{ShMem}) into Taylor series in powers of $\tau$. In zero order we get an equality $\Psi=\Psi$.
In first order we obtain an expected trivial ``infinitesimal" result $\Upsilon_0 (\Psi) =
P_{\Psi}(J(\Psi))$. First non-trivial result is an expression for $\Upsilon_1 (\Psi)$:
\begin{eqnarray}\label{Upsilon_1}
\Upsilon_1 (\Psi)&=&{1 \over 2} \{P_{\Psi}[(D_{\Psi}J({\Psi}))J({\Psi})] -
[D_{\Psi}(P_{\Psi}J(\Psi))](P_{\Psi}J(\Psi))\} + \nonumber \\ && {1 \over 2} [D^2_{\Psi}{\bf
P}_{\Omega}(\Psi)](J(\Psi),J(\Psi)).
\end{eqnarray}
In this order we obtain the first short-memory approximation:
\begin{eqnarray}\label{ShM}
{d\Psi \over dt}&=&P_{\Psi}(J(\Psi))+ \nonumber \\ && {\tau \over 2}
\{P_{\Psi}[(D_{\Psi}J({\Psi}))J({\Psi})] - [D_{\Psi}(P_{\Psi}J(\Psi))](P_{\Psi}J(\Psi))\} + \nonumber
\\ && {\tau \over 2} [D^2_{\Psi}{\bf P}_{\Omega}(\Psi)](J(\Psi),J(\Psi)).
\end{eqnarray}
If $\ker P_{\Psi}$ does not depend of $\Psi$, then it is possible to choose such a coordinate system
on $\Omega$ where linear operator $P_{\Psi}$ does not depend of $\Psi$. Then the last term in Eqs.
(\ref{Upsilon_1}), (\ref{ShM})  vanishes. It is the case of quasiequilibrium manifolds, for example
\cite{GKIOeNONNEWT2001,GKOeTPRE2001}.

Various physical examples of application of these formulae with quasiequilibrium manifolds are
presented in refs. \cite{GKIOeNONNEWT2001,GKOeTPRE2001,GKMex2001,GKPRE02,KTGOePhA2003}.

The theory of short memory and coarse-graining in the form given by Eqs. (\ref{ShMem}),
(\ref{Upsilon_1}), (\ref{ShM}) has one free parameter: the memory time $\tau (\Psi)$. The next step
is development of the theory without such free parameters \cite{GKGeoNeo,Plenka}.

The first attempt to formalize the short memory approximation and coarse-graining on the base of the
matching equation (\ref{ShMem}) was made by Lewis \cite{Lew}, but he expanded only the right hand
side of Eq. (\ref{ShMem}), and the result was not a solution of this equation. Very recently the
short memory approximation becomes more popular \cite{Cho,Calz}.

\subsection{The theorem about entropy overproduction in the short memory approximation}

The short memory approximation (\ref{ShM}) has one important property: it increase the entropy
production for the thermodynamic projectors $P_{\Psi}$: if for any vector field the field of
projectors $P_{\Psi}$ preserves dissipation, then for any vector field the short memory approximation
increases dissipation (and it strictly increases dissipation, if the vector field is not tangent to
ansatz manifold $\Omega$).

This {\it theorem about entropy overproduction} can be formulate as a expression for entropy
production.

{\bf Theorem 2.} {\it Let $P_{\Psi}$ be the field of thermodynamic projectors. Then, due to the short
memory approximation (\ref{ShM}) for $\Psi \in \Omega$}
\begin{equation}\label{EProSM}
{dS \over dt}= \sigma_0(\Psi) + {\tau \over 2}\sigma_1(\Psi) + {\tau \over 2} \langle \Delta(\Psi)
|\Delta(\Psi) \rangle_{\Psi},
\end{equation}
{\it where $\Delta$ is defect of invariance: $\Delta = J(\Psi)-P_{\Psi}(J(\Psi)),$ }
\begin{eqnarray}
&&\sigma_0(\Psi)=\left. {d S({\bf T}_{\tau}(\Psi)) \over d \tau}\right|
_{\tau=0}=(D_{\Psi}S(\Psi))(J(\Psi)); \nonumber \\ &&\sigma_1(\Psi)= \left. {d^2 S({\bf
T}_{\tau}(\Psi)) \over d \tau^2}\right| _{\tau=0} = (D_{\Psi}S(\Psi)) [(D_{\Psi}J(\Psi))(J(\Psi))] -
\langle J(\Psi) | J(\Psi) \rangle_{\Psi}.
\end{eqnarray}
($\sigma_0(\Psi)$ is entropy production due to initial system at a point $\Psi$.) The geometrical
proof of this formula (\ref{EProSM}) is simple. The matching condition (\ref{ShMem}) represents a
motion in according to a short memory approximation as two steps: (i) the motion along trajectories
of initial system during time $\tau$, and (ii) the projection onto manifold $\Omega$ by means of the
operator ${\bf P}_{\Omega}$ (\ref{glopro}), (\ref{ProPre}). The entropy increment on the first step
is $$S({\bf T}_{\tau}(\Psi))-S(\Psi) = \tau \sigma_0(\Psi) + {\tau^2 \over 2}\sigma_1(\Psi) +
o(\tau^2).$$ To calculate the entropy increment on the projection step with accuracy $\tau^2$ we need
to calculate the motion with accuracy $\tau^1$ only, because the point $\Psi \in \Omega$ is the point
of entropy maximum on its preimage  ${\bf P}_{\Omega}^{-1} (\Psi)$ (here it is crucial that
$P_{\Psi}$ is the field of thermodynamic projectors with the property ${\bf A}$ (\ref{condA})). In
the first order ${\bf T}_{\tau}(\Psi)-{\bf P}_{\Omega}({\bf T}_{\tau}(\Psi)) = \tau \Delta(\Psi)$ and
the last term in Eq. (\ref{EProSM}) is just the Taylor formula for $S$ of the second order. $\square$

First two terms in Eq. (\ref{EProSM}) give the average entropy production by the initial system in
time $\tau$ (up to the second order in $\tau$). The third term is always non-negative, and is zero
only for zero defect of invariance. In this sense Theorem 2 is {\it the theorem about entropy
overproduction}. The following Corollary gives an obvious, but physically important consequence of
this theorem.

{\bf Corollary 3.} {\it Let the initial system (\ref{sys}) be conservative: $dS/dt =
(D_{\Psi}S(\Psi))(J(\Psi)) \equiv 0$. Then for the short memory approximation (\ref{ShM}) entropy
production is non-negative:}
\begin{equation}\label{EProCon}
{dS \over dt} = {\tau \over 2} \langle \Delta(\Psi) |\Delta(\Psi) \rangle_{\Psi} \geq 0,
\end{equation}
{\it and $dS/dt=0$ if and only if the vector field $J(\Psi)$ is tangent to ansatz manifold.}
$\square$

The short-memory approximation equipped by the thermodynamic projector gives us the simplest way to
introduce dissipation into a conservative system.

\section{The art of ansatz: Multi-peak polyhedrons in kinetic systems with instabilities}

\subsection{Two-peak approximations}

The thermodynamic projector guarantees the thermodynamic consistence of ansatz, and post-processing
gives both the estimations of the error and correction for the solution. So, the main requirement to
an ansatz now is: to capture the essence of the phenomenon. This is {\it the art of ansatz}. Is it
possible to formalize this art? In this subsection we discuss two special ansatz which are known for
several decades and mysteriously are at the same time simplest and reliable nonperturbative
approximations in the domains of their application. The requested formalization seems to be possible,
at least, partially.

\subsubsection{Tamm--Mott-Smith approximation for kinetics of shock waves}

Shock waves in gas flows are important from practical, as well, as from theoretical points of view.
Some integral parameters of the shock wave front can be obtained by gas dynamics equations with
additional thermodynamic relations, for weak shocks the hydrodynamic approach can give the shock
front structure too \cite{LaLi}. For strong shocks it is necessary to use the kinetic representation,
for rarefied gases the Boltzmann kinetic equation gives the framework for studying the structure of
strong shocks \cite{Cercignani}. This equation describes the dynamics of the one-particle
distribution function $f(\vv, \xx),$ where $\vv$ is the vector of particle velocity, and $\xx$ is the
particle position in space. One of the common ways to use the Boltzmann equation in physics away from
exact solutions and perturbation expansions  consists of three steps:
\begin{enumerate}
\item{Construction of a specific ansatz for the distribution function for a given physical problem;}
\item{Projection of the Boltzmann equation on the ansatz;}
\item{Estimation and correction of the ansatz (optional).}
\end{enumerate}
The first and, at the same time, the  most successful ansatz for the distribution function in the
shock layer was invented in the middle of the XX century. It is the bimodal Tamm--Mott-Smith
approximation (see, for example, the book \cite{Cercignani}):
\begin{equation}\label{TMS}
f(\vv, \xx)=f_{\rm TMS}(\vv, z) = a_-(z)f_-(\vv)+a_+(z)f_+(\vv),
\end{equation}
where $z$ is the space coordinate in the direction of the shock wave motion, $f_\pm(\vv)$ are the
downstream and the upstream Maxwellian distributions, respectively.

Direct molecular dynamics simulation for the Lennard-Jones gas shows good {\it quantitative}
agreement of the Tamm--Mott-Smith ansatz (\ref{TMS}) with the simulated velocity distribution
functions in the shock fronts for a wide  range of Mach number (between 1 and 8.19) \cite{Zhakh}. For
different points in the shock front the bimodal approximation (\ref{TMS}) of the simulated velocity
distribution function has appropriate accuracy, but the question about approximation of the
$a_\pm(z)$ remained open in the paper \cite{Zhakh}, because  the authors of this paper had ``no way
to decide which of the equations proposed in the literature yields better results".

The thermodynamic projector gives the unique thermodynamically consistent equation for the
Tamm--Mott-Smith approximation (\ref{TMS}) \cite{GK1} (in our paper \cite{GK1} we have used only {\it
the property } ${\bf A}$, but for this one-dimensional ansatz it was sufficient for uniqueness of the
projector). These equations have a simple form for the variables: $$n(z)= \int f_{\rm TMS}(\vv, z)
d^3 \vv; \: s(z)=-k_{\rm B}\int f_{\rm TMS}(\vv, z)\ln f_{\rm TMS}(\vv, z) d^3 \vv.$$ The particles
density $n(z)$ is linear function of $a_\pm(z)$. The entropy density $s$ is a more complicated
function of $a_\pm$, but there are simple expansions both for weak and for strong shocks
\cite{GK1,Lampis}.

The equations for $n(z,t), \: s(z,t)$ in the Tamm--Mott-Smith approximation have the form:
\begin{equation}\label{TMSeq}
{\partial s \over \partial t} + {\partial j_s \over \partial z} = \sigma,  \: {\partial n \over
\partial t} + {\partial j_n \over
\partial z}=0,
\end{equation}
where $$j_s(z)=-k_{\rm B}\int v_z f_{\rm TMS}(\vv, z)\ln f_{\rm TMS}(\vv, z) d^3 \vv, \: j_n(z)= \int
v_z f_{\rm TMS}(\vv, z) d^3 \vv,$$ and $\sigma$ is the Boltzmann density of entropy production for
the TMS distribution (\ref{TMS}): $$\sigma=-k_{\rm B}\int J(f_{\rm TMS})(\vv, z)\ln f_{\rm TMS}(\vv,
z) d^3 \vv,$$ where $J(f)$ is the Boltzmann collision integral.

Equations (\ref{TMSeq}) were first introduced  by M. Lampis \cite{Lampis} in the ad hoc manner.
Direct numerical simulation demonstrated that all other known equations for the Tamm--Mott-Smith
ansatz violate the second law  \cite{Naka}.

\subsubsection{Langer--Bar-on--Miller approximation for spinodal decomposition}

The spinodal decomposition is the initial stage of a phase separation in thermodynamically unstable
solid solution. It requires no activation energy (unstable does not mean metastable). The order
parameter is the composition variable (concentration $c$ of one of components, for example). Hence,
the rate of the spinodal decomposition is limited by diffusion processes.

The process of spinodal decomposition was described quantitatively in the paper  \cite{Spinodal}.
This model consists of two coupled equations: for the single-point distribution function of
fluctuations, and for the pair correlation function. The fluctuation $u(\rr)=c(\rr)-c_0$ is deviation
of the concentration $c$ from the average concentration $c_0$. The time evolution of the single-point
distribution density of fluctuation, $\rho_1(u)$ is described by the one-dimensional Fokker-Planck
equation:
\begin{equation}\label{FPESpin}
{\partial \rho_1 \over \partial t}= M {\partial  \over
\partial u} \left(  \rho_1{\partial  F(u)  \over
\partial u} + k_{\rm B}T b { \partial \rho_1 \over \partial
u}\right) ,
\end{equation}
where $b$ is a constant, $F(u)$ is a mean-field free energy which depends on the value of $u$, on the
whole function $\rho_1$ (because $F(u)$ includes some averages in the mean field approximation), and
on the two-point correlation function (because it depends on average square of $\nabla u(\rr)$). The
assumption
\begin{equation}\label{Corr2}
\rho_2[u(\rr),u(\rr_0)]\cong \rho_1[u(\rr)]\rho_1[u(\rr_0)]\{1+ \gamma (|\rr-\rr_0|)u(\rr)u(\rr_0) \}
\end{equation}
allows to truncate the infinite  chain of equations for all correlation functions, and to write the
equation for the two-point correlation function. Details are presented in the paper \cite{Spinodal}.

The mean-field free energy function $F(u)$ is non-stationary and may be non-convex. Thus, the
one-peak representations for $\rho_1(u)$ are far from a physical sense, but it is possible to try the
two-peak ansatz:
\begin{equation}\label{TwoPeakSpi}
\rho_1(u)=a_1 G_{\sigma}(u-\varsigma_1)+a_2 G_{\sigma}(u+\varsigma_2),
\end{equation}
where $a_1=\varsigma_2/(\varsigma_1+\varsigma_2)$, $a_2=\varsigma_1/(\varsigma_1+\varsigma_2)$
(because obvious normalization conditions), and $G_{\sigma}(u)$ is the Gaussian distribution:
$G_{\sigma}(u)= \frac{1}{\sigma\sqrt{2\pi}}\exp\left(-\frac{u^2}{2\sigma^2}\right).$

The systematic use of this two-peak ansatz (\ref{TwoPeakSpi}) allowed to get the satisfactory
quantitative description for some features of spinodal decomposition. The authors of the paper
\cite{Spinodal} mentioned that the present computational scheme does appear to be accurate enough to
justify its use in the study of realistic metallurgical systems. Instead of thermodynamic projector
which was not known in 1975, they used the projection onto the first three non-trivial moments
($\langle u^2 \rangle, \: \langle u^3 \rangle, \: \langle u^4 \rangle$).

The Langer--Bar-on--Miller bimodal ansatz has a long history of criticizing and comparison with
experiments and other calculations. During 10 years after this publication there were some attempts
to criticize and improve this theory. Nevertheless, at 1985 this theory was called ``the most
successful ``early-time" theory yet available" \cite{Grant}. In this paper the nonlinear
Langer--Bar-on--Miller theory was again criticized as non-systematic, ``since there is no smallness
parameter"\footnote{This explicit belief in small parameters and Taylor expansion remains widespread
in spite of many well-known computational algorithms that use no explicit small parameters both in
computational mathematics and mathematical physics. It seems strange after success of famous
KAM-theory, for example.}. Nevertheless, 20 years of various attempts to improve this theory of
bimodal ansatz were resumed in the paper \cite{Kum}: ``There have been many theories that attempt to
incorporate nonlinear effects in the description of the spinodal decomposition process. The most
successful of these was devised by Langer, Baron, and Miller". This situation reminds the situation
with TMS ansatz for shock waves.

More recently, applications of Monte Carlo methods to the self-consistent calculation of a
Ginzburg-Landau free energy functional for Lennard-Jones systems in three dimensions are discussed
\cite{Gra}. It was demonstrated that the parameters in the coarse-grained free energy can be
extracted from a multivariate distribution of energies and particle densities. Histograms  of
calculated unimodal and bimodal density distributions are presented.

\subsection{Multi-peak ansatz and mean-field theory of molecular individualism}

\subsubsection{Two-peak approximation for polymer stretching in flow
and explosion of the Gaussian manifold}

We shall consider the simplest case of dilute polymer solutions represented by dumbbell models. The
dumbbell model reflects the two features of real--world macromolecules to be orientable and
stretchable by a flowing solvent \cite{Bird}.

Let us consider the simplest one-dimensional kinetic equation for the configuration distribution
function $\Psi(q,t)$, where $q$ is the reduced vector connecting the beads of the dumbbell. This
equation is slightly different from the usual Fokker-Planck equation. It is nonlinear, because of the
dependence of potential energy $U$ on the moment $M_{2}[\Psi]=\int q^{2}\Psi(q) dq$. This dependence
allows us to get the exact quasiequilibrium equations on $M_{2}$, but this equations are not solving
the problem: this quasiequilibrium manifold may become unstable when the flow is present \cite{IK00}.
Here is this model:
\begin{equation}\label{530}
\partial_{t}\Psi=-\partial_{q}\{\alpha(t)q\Psi\}+\frac{1}{2}\partial^{2}_{q}\Psi.
\end{equation}
Here
\begin{equation}\label{531}
\alpha(t)=\kappa(t)-\frac{1}{2}f(M_{2}(t)),
\end{equation}
$\kappa(t)$ is the given time-independent velocity gradient, $t$ is the reduced time, and the
function $-fq$ is the reduced spring force. Function $f$ may depend on the second moment of the
distribution function $M_{2}=\int q^{2}\Psi(q,t)dq$. In particular, the case $f\equiv1$ corresponds
to the linear Hookean spring, while $f=[1-M_{2}(t)/b]^{-1}$ corresponds to the self-consistent finite
extension nonlinear elastic spring (the FENE-P model, first introduced in \cite{FENEP}). The second
moment $M_{2}$ occurs in the FENE-P force $f$ as the result of the pre-averaging approximation to the
original FENE model (with nonlinear spring force $f=[1-q^{2}/b]^{-1}$). The parameter $b$ changes the
characteristics of the force law from Hookean at small extensions to a confining force for
$q^{2}\rightarrow b$. Parameter $b$ is roughly equal to the number of monomer units represented by
the dumbell and should therefore be a large number. In the limit $b\rightarrow\infty$, the Hookean
spring is recovered. Recently, it has been demonstrated that FENE-P model appears as first
approximation within a systematic self-confident expansion of nonlinear forces
\cite{GKIOeNONNEWT2001}.

Equation (\ref{530}) describes an ensemble of non-interacting dumbells subject to a
pseudo-elongational flow with fixed kinematics. As is well known, the Gaussian distribution function,
\begin{equation}\label{532}
\Psi^{G}(M_{2})=\frac{1}{\sqrt{2\pi M_{2}}}\exp\left[-\frac{q^{2}}{2M_{2}}\right],
\end{equation}
solves equation (\ref{530}) provided the second moment $M_{2}$ satisfies
\begin{equation}\label{533}
\frac{dM_{2}}{dt}=1+2\alpha(t)M_{2}.
\end{equation}
Solution (\ref{532}) and (\ref{533}) is the valid macroscopic description if all other solutions of
the equation (\ref{530}) are rapidly attracted to the family of Gaussian distributions (\ref{532}).
In other words \cite{GKTTSP94}, the special solution (\ref{532}) and (\ref{533}) is the macroscopic
description if equation (\ref{532}) is the stable invariant manifold of the kinetic equation
(\ref{530}). If not, then the Gaussian solution is just a member of the family of solutions, and
equation (\ref{533}) has no meaning of the macroscopic equation. Thus, the complete answer to the
question of validity of the equation (\ref{533}) as the macroscopic equation requires a study of
dynamics in the neighborhood of the manifold (\ref{532}). Because of the simplicity of the model
(\ref{530}), this is possible to a satisfactory level even for $M_{2}$-dependent spring forces.

In the paper \cite{IK00} it was shown, that there is a possibility of ``explosion" of the Gaussian
manifold: with the small initial deviation from it, the solutions of the equation (\ref{530}) are
very fast going far from, and then slowly come back to the stationary point which is located on the
Gaussian manifold. The distribution function $\Psi$ is stretched fast, but looses the Gaussian form,
and after that the Gaussian form recovers slowly with the new value of $M_{2}$. Let us describe
briefly the results of \cite{IK00}.

Let $M_{2n}=\int q^{2n}\Psi dq$ denote the even moments (odd moments vanish by symmetry). We consider
deviations $\mu_{2n}=M_{2n}-M_{2n}^{\rm G}$, where $M_{2n}^{\rm G}=\int q^{2n} \Psi^{\rm G}dq$ are
moments of the Gaussian distribution function (\ref{532}). Let $\Psi(q,t_0)$ be the initial condition
to the Eq.\ (\ref{530}) at time $t=t_0$. Introducing functions,
\begin{equation}
\label{result0} p_{2n}(t,t_0)=\exp\left[4n\int_{t_0}^{t}\alpha(t')dt'\right],
\end{equation}
where $t\ge t_0$, and $2n \ge 4$, the {\it exact} time evolution of the deviations $\mu_{2n}$ for
$2n\ge 4$ reads
\begin{equation} \label{result1}
    \mu_4(t)=p_4(t,t_0)\mu_4(t_0),
\end{equation}
and
\begin{equation} \label{result2}
    \mu_{2n}(t)=\left[ \mu_{2n}(t_0) + 2n(4n-1)\int_{t_0}^t
    \mu_{2n-2}(t')p_{2n}^{-1}(t',t_0)dt' \right] p_{2n}(t,t_0),
\end{equation}
for $2n\ge 6$. Equations (\ref{result0}), (\ref{result1}) and (\ref{result2}) describe evolution near
the Gaussian solution for arbitrary initial condition $\Psi(q,t_0)$. Notice that explicit evaluation
of the integral in the Eq.\ (\ref{result0}) requires solution to the moment equation (\ref{533})
which is not available in the analytical form for the FENE-P model.

It is straightforward to conclude that any solution with a non-Gaussian initial condition converges
to the Gaussian solution asymptotically as $t\to\infty$ if

\begin{equation}
\label{result3} \lim_{t\to\infty}\int_{t_0}^t\alpha(t')dt'<0.
\end{equation}
However, even if this asymptotic condition is met, deviations from the Gaussian solution may survive
for considerable {\it finite} times. For example, if for some finite time $T$, the integral in the
Eq.\ (\ref{result0}) is estimated as $\int_{t_0}^t\alpha(t')dt'>\alpha (t-t_0)$, $\alpha>0$, $t\le
T$, then the Gaussian solution becomes exponentially unstable during this time interval. If this is
the case, the moment equation (\ref{533}) cannot be regarded as the macroscopic equation. Let us
consider specific examples.

For the Hookean spring ($f\equiv 1$) under a constant elongation ($\kappa={\rm const}$), the Gaussian
solution is exponentially stable for $\kappa<0.5$, and it becomes exponentially unstable for
$\kappa>0.5$. The exponential instability in this case is accompanied by the well known breakdown of
the solution to the Eq.\ (\ref{533}) due to infinite stretching of the dumbbell. The situation is
much more interesting for the FENE-P model because this nonlinear spring force does not allow the
infinite stretching of the dumbbell.

\begin{figure}[t]
\centering{
\includegraphics[width=130mm,height=120mm]{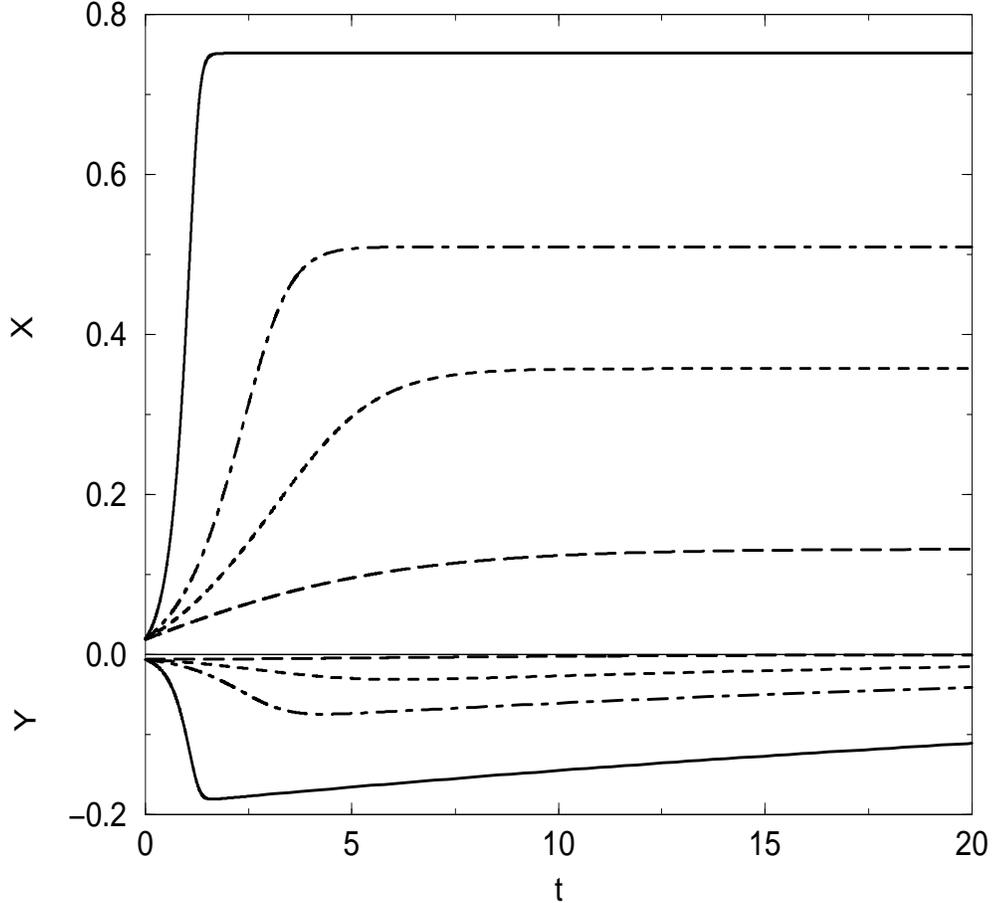}
\caption{Deviations of reduced moments from the Gaussian solution as a function of reduced time $t$
in pseudo-elongation flow for the FENE-P model. Upper part: Reduced second moment $X=M_2/b$. Lower
part: Reduced deviation of fourth moment from Gaussian solution $Y=-\mu_4^{1/2}/b$. Solid:
$\kappa=2$, dash-dot: $\kappa=1$, dash: $\kappa=0.75$, long dash: $\kappa=0.5$. (The figure from the
paper \cite{IK00}, computed by P. Ilg.)} \label{EPJ713_fig}}
\end{figure}

Eqs.\ (\ref{533}) and (\ref{result1}) were integrated  by the 5-th order Runge-Kutta method with
adaptive time step. The FENE-P parameter $b$ was set equal to 50. The initial condition was
$\Psi(q,0)=C(1-q^2/b)^{b/2}$, where $C$ is the normalization (the equilibrium of the FENE model,
notoriously close to the FENE-P equilibrium \cite{Herrchen}). For this initial condition, in
particular, $\mu_4(0)=-6b^2/[(b+3)^2(b+5)]$ which is about 4$\%$ of the value of $M_4$ in the
Gaussian equilibrium for $b=50$. In Fig.~\ref{EPJ713_fig} we demonstrate deviation $\mu_4(t)$ as a
function of time for several values of the flow. Function $M_2(t)$ is also given for comparison. For
small enough $\kappa$ we find an adiabatic regime, that is $\mu_4$ relaxes exponentially to zero. For
stronger flows, we observe an initial {\it fast runaway} from the invariant manifold with $|\mu_4|$
growing over three orders of magnitude compared to its initial value. After the maximum deviation has
been reached, $\mu_4$ relaxes to zero. This relaxation is exponential as soon as the solution to Eq.\
(\ref{533}) approaches the steady state. However, the time constant for this exponential relaxation
$|\alpha_{\infty}|$ is very small. Specifically, for large $\kappa$,
\begin{equation}
\label{alpha_lim} \alpha_{\infty}=\lim_{t\to\infty}\alpha(t)=-\frac{1}{2b}+O(\kappa^{-1}).
\end{equation}
Thus, the steady state solution is unique and Gaussian but the stronger is the flow, the larger is
the initial runaway from the Gaussian solution, while the return to it thereafter becomes
flow-independent. Our observation demonstrates that, though the stability condition (\ref{result3})
is met, {\it significant deviations from the Gaussian solution persist over the times when the
solution of Eq.}\ (\ref{533}) {\it is already reasonably close to the stationary state.} If we accept
the usually quoted physically reasonable minimal value of parameter $b$ of the order $20$ then the
minimal relaxation time is of order $40$ in the reduced time units of Fig.~\ref{EPJ713_fig}. We
should also stress that the two limits, $\kappa\to\infty$ and $b\to\infty$, are not commutative, thus
it is not surprising that the estimation (\ref{alpha_lim}) does not reduce to the above mentioned
Hookean result as $b\to\infty$. Finally, peculiarities of convergence to the Gaussian solution are
even furthered if we consider more complicated (in particular, oscillating) flows $\kappa(t)$.
Further numerical experiments are presented in \cite{Lisa}. The statistics of FENE-P solutions with
random strains was studied recently by J.-L. Thiffeault \cite{JLT}

In accordance with \cite{Legendre} the ansatz for $\Psi$ can be suggested in the following form:
\begin{equation}\label{Anz}
\Psi^{An}(\{\sigma,\varsigma\},q)=
\frac{1}{2\sigma\sqrt{2\pi}}\left(e^{-\frac{(q+\varsigma)^{2}}{2\sigma^{2}}}+
e^{-\frac{(q-\varsigma)^{2}}{2\sigma^{2}}}\right).
\end{equation}
Natural inner coordinates on this manifold are $\sigma$ and $\varsigma$. Note, that now
$\sigma^{2}\neq M_{2}$. The value $\sigma^{2}$ is a dispersion of one of the Gaussian summands in
(\ref{Anz}),
\begin{eqnarray*}
M_{2}(\Psi^{An}(\{\sigma,\varsigma\},q))=\sigma^{2}+\varsigma^{2}.
\end{eqnarray*}
To build the thermodynamic projector on the manifold (\ref{Anz}), the thermodynamic Lyapunov function
is necessary. It is necessary to emphasize, that equations (\ref{530}) are nonlinear. For such
equations, the arbitrarity in the choice of the thermodynamic Lyapunov function is much smaller than
for the linear Fokker Planck equation. Nevertheless, such a function exists. It is the free energy
\begin{equation}\label{Free}
F=U(M_{2}[\Psi])-TS[\Psi],
\end{equation}
where
\begin{eqnarray*}
S[\Psi]=-\int\Psi(\ln\Psi-1)dq,
\end{eqnarray*}
$U(M_{2}[\Psi])$ is the potential energy in the mean field approximation, $T$ is the temperature
(further we assume that $T=1$).

 Note, that Kullback-form entropy
$S_{k}=-\int\Psi\ln\left(\frac{\Psi}{\Psi^{*}}\right)$ also has the form $S_{k}=-F/T$:
\begin{eqnarray*}
\Psi^{*}=\exp(-U),\\ S_{k}[\Psi]=-\langle U\rangle-\int\Psi\ln\Psi dq.
\end{eqnarray*}
If $U(M_{2}[\Psi])$ in the mean field approximation is the convex function of $M_{2}$, then the free
energy (\ref{Free}) is the convex functional too.

For the FENE-P model $U=-\ln[1-M_{2}/b]$.

In accordance to the thermodynamics the vector $I$ of flow of $\Psi$ must be proportional to the
gradient of the corresponding chemical potential $\mu$:
\begin{equation}\label{Flux}
I=-B(\Psi)\nabla_{q}\mu,
\end{equation}
where $\mu=\frac{\delta F}{\delta\Psi}$, $B\geq0$. From the equation (\ref{Free}) it follows, that
\begin{eqnarray}\label{muflux}
\mu=\frac{d U(M_{2})}{d M_{2}}\cdot q^{2}+\ln\Psi\nonumber\\ I=-B(\Psi)\left[2\frac{dU}{dM_{2}}\cdot
q+\Psi^{-1}\nabla_{q}\Psi\right].
\end{eqnarray}
If we suppose here $B=\frac{D}{2}\Psi$, then we get
\begin{eqnarray}\label{TDeq}
I=-D\left[\frac{dU}{dM_{2}}\cdot q\Psi+\frac{1}{2}\nabla_{q}\Psi\right]\nonumber\\
\frac{\partial\Psi}{\partial t}={\rm div}_{q}I=D\frac{d U(M_{2})}{d
M_{2}}\partial_{q}(q\Psi)+\frac{D}{2}\partial^{2}q\Psi,
\end{eqnarray}
When $D=1$ this equations coincide with (\ref{530}) in the absence of the flow: due to equation
(\ref{TDeq}) $dF/dt\leq0$.

Let us construct the thermodynamic projector with the help of the thermodynamic Lyapunov function $F$
(\ref{Free}). Corresponding entropic scalar product in the point $\Psi$ has the form
\begin{equation}\label{Scal}
\left.\langle f|g\rangle_{\Psi}=\frac{d^{2}U}{dM_{2}^{2}}\right|_{M_{2}=M_{2}[\Psi]}\cdot\int
q^{2}f(q)dq\cdot\int q^{2}g(q)dq+\int\frac{f(q)g(q)}{\Psi(q)}dq
\end{equation}
During the investigation of the ansatz (\ref{Anz}) the scalar product (\ref{Scal}), constructed for
the corresponding point of the Gaussian manifold with $M_{2}=\sigma^{2}$, will be used. It will let
us to investigate the neighborhood of the Gaussian manifold (and to get all the results in the
analytical form):
\begin{equation}\label{ScalG}
\left.\langle f|g\rangle_{\sigma^{2}}=\frac{d^{2}U}{dM_{2}^{2}}\right|_{M_{2}=\sigma^{2}}\cdot\int
q^{2}f(q)dq\cdot\int q^{2}g(q)dq+\sigma\sqrt{2\pi}\int e^{\frac{q^{2}}{2\sigma^{2}}}f(q)g(q)dq
\end{equation}
Also we will need to know the functional $DF$ in the point of Gaussian manifold:
\begin{equation}\label{Prod}
\left.DF_{\sigma^{2}}(f)=\left(\frac{d U(M_{2})}{dM_{2}}\right|_{M_{2}=\sigma^{2}}
-\frac{1}{2\sigma^{2}}\right)\int q^{2}f(q)dq,
\end{equation}
\noindent (with the condition $\int f(q)dq=0$). The point
\begin{eqnarray*}
\left.\frac{d U(M_{2})}{dM_{2}}\right|_{M_{2}=\sigma^{2}}=\frac{1}{2\sigma^{2}},
\end{eqnarray*}
corresponds to the equilibrium.

The tangent space to the manifold (\ref{Anz}) is spanned by the vectors
\begin{eqnarray}\label{basis}
&&f_{\sigma}=\frac{\partial\Psi^{An}}{\partial(\sigma^{2})}; \:
f_{\varsigma}=\frac{\partial\Psi^{An}}{\partial(\varsigma^{2})};\nonumber\\
f_{\sigma}&=&\frac{1}{4\sigma^{3}\sqrt{2\pi}}\left[e^{-\frac{(q+\varsigma)^{2}}{2\sigma^{2}}}
\frac{(q+\varsigma)^{2}-\sigma^{2}}{\sigma^{2}}+e^{-\frac{(q-\varsigma)^{2}}{2\sigma^{2}}}
\frac{(q-\varsigma)^{2}-\sigma^{2}}{\sigma^{2}} \right];\\
f_{\varsigma}&=&\frac{1}{4\sigma^{2}\varsigma\sqrt{2\pi}}\left[-e^{-\frac{(q+\varsigma)^{2}}{2\sigma^{2}}}
\frac{q+\varsigma}{\sigma}+e^{-\frac{(q-\varsigma)^{2}}{2\sigma^{2}}} \frac{(q-\varsigma)}{\sigma}
\right];\nonumber
\end{eqnarray}
The Gaussian entropy (free energy) production in the directions $f_{\sigma}$ and $f_{\varsigma}$
(\ref{Prod}) has a very simple form:
\begin{eqnarray}\label{Fpro}
\left.DF_{\sigma^{2}}(f_{\varsigma})=DF_{\sigma^{2}}(f_{\sigma})=\frac{d
U(M_{2})}{dM_{2}}\right|_{M_{2}=\sigma^{2}}-\frac{1}{2\sigma^{2}}.
\end{eqnarray}
The linear subspace $\ker DF_{\sigma^{2}}$ in $lin\{f_{\sigma},f_{\varsigma}\}$ is spanned by the
vector $f_{\varsigma}-f_{\sigma}$.

Let us have the given vector field $d\Psi/dt=J(\Psi)$ in the point $\Psi(\{\sigma,\varsigma\})$. We
need to build the projection of $J$ onto the tangent space $T_{\sigma,\varsigma}$ in the point
$\Psi(\{\sigma,\varsigma\})$:
\begin{equation}\label{Prosigma}
P^{th}_{\sigma,\varsigma}(J)=\varphi_{\sigma}f_{\sigma}+\varphi_{\varsigma}f_{\varsigma}.
\end{equation}
This equation means, that the equations for $\sigma^{2}$ and $\varsigma^{2}$ will have the form
\begin{equation}\label{eqsigma}
\frac{d\sigma^{2}}{dt}=\varphi_{\sigma};\:\: \frac{d\varsigma^{2}}{dt}=\varphi_{\varsigma}
\end{equation}
Projection $(\varphi_{\sigma},\varphi_{\varsigma})$ can be found from the following two equations:
\begin{eqnarray}\label{psieq}
\varphi_{\sigma}+\varphi_{\varsigma}=\int q^{2}J(\Psi)(q)dq\nonumber;\\ \langle
\varphi_{\sigma}f_{\sigma}+\varphi_{\varsigma}f_{\varsigma}|f_{\sigma}-f_{\varsigma}\rangle_{\sigma^{2}}
=\langle J(\Psi)|f_{\sigma}-f_{\varsigma}\rangle_{\sigma^{2}},
\end{eqnarray}
where $\langle f|g\rangle_{\sigma^{2}}=\langle J(\Psi)|f_{\sigma}-f_{\varsigma}\rangle_{\sigma^{2}}$,
(\ref{Scal}). First equation of (\ref{psieq}) means, that the time derivative $dM_{2}/dt$ is the same
for the initial and the reduced equations. Due to the formula for the dissipation of the free energy
(\ref{Prod}), this equality is equivalent to the persistence of the dissipation in the neighborhood
of the Gaussian manifold. Indeed, in according to (\ref{Prod}) $dF/dt=A(\sigma^{2})\int q^2
J(\Psi)(q)dq= A(\sigma^{2}) dM_2/dt$, where $A(\sigma^{2})$ does not depend of $J$. On the other
hand, time derivative of $M_2$ due to projected equation (\ref{eqsigma}) is
$\varphi_{\sigma}+\varphi_{\varsigma}$, because $M_2=\sigma^2+\varsigma^2$.

The second equation in (\ref{psieq}) means, that $J$ is projected orthogonally on $\ker DS\bigcap
T_{\sigma,\varsigma}$. Let us use the orthogonality with respect to the entropic scalar product
(\ref{ScalG}). The solution of equations (\ref{psieq}) has the form
\begin{eqnarray}\label{projphi}
\frac{d\sigma^{2}}{dt}=\varphi_{\sigma}=\frac{\langle
J|f_{\sigma}-f_{\varsigma}\rangle_{\sigma^{2}}+M_{2}(J)(\langle
f_{\varsigma}|f_{\varsigma}\rangle_{\sigma^{2}}-\langle
f_{\sigma}|f_{\varsigma}\rangle_{\sigma^{2}})}{\langle
f_{\sigma}-f_{\varsigma}|f_{\sigma}-f_{\varsigma}\rangle_{\sigma^{2}}}\nonumber,\\\\
\frac{d\varsigma^{2}}{dt}=\varphi_{\varsigma}=\frac{-\langle
J|f_{\sigma}-f_{\varsigma}\rangle_{\sigma^{2}}+M_{2}(J)(\langle
f_{\sigma}|f_{\sigma}\rangle_{\sigma^{2}}-\langle
f_{\sigma}|f_{\varsigma}\rangle_{\sigma^{2}})}{\langle
f_{\sigma}-f_{\varsigma}|f_{\sigma}-f_{\varsigma}\rangle_{\sigma^{2}}}\nonumber,
\end{eqnarray}
where $J=J(\Psi)$, $M_{2}(J)=\int q^{2}J(\Psi)dq$.

\begin{figure}[t]
\begin{centering}
\includegraphics[width=160mm, height=127mm]{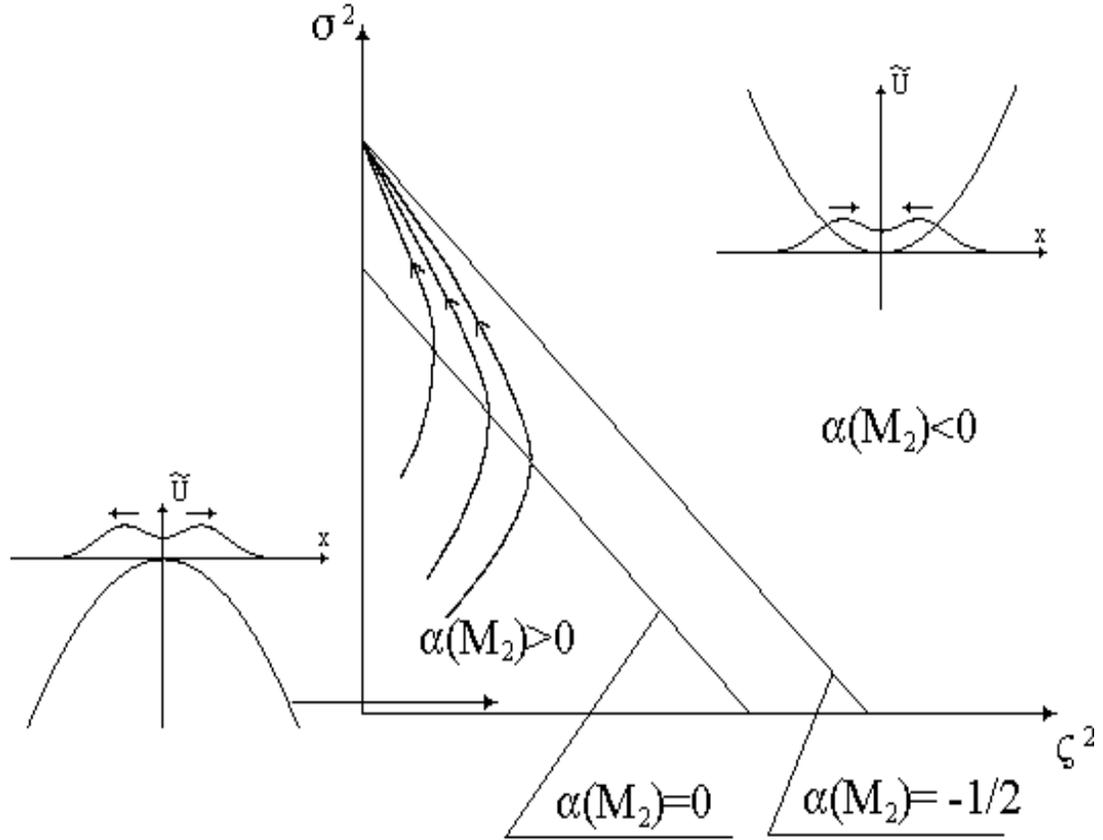}
\caption {Phase trajectories for the two-peak approximation, FENE-P model. The vertical axis
($\varsigma=0$) corresponds to the Gaussian manifold. The triangle with $\alpha(M_2)>0$ is the domain
of exponential instability. }
    \label{figFENEP}
\end{centering}
\end{figure}

\begin{figure}[t]
\begin{centering}
\includegraphics[width=160mm, height=90mm]{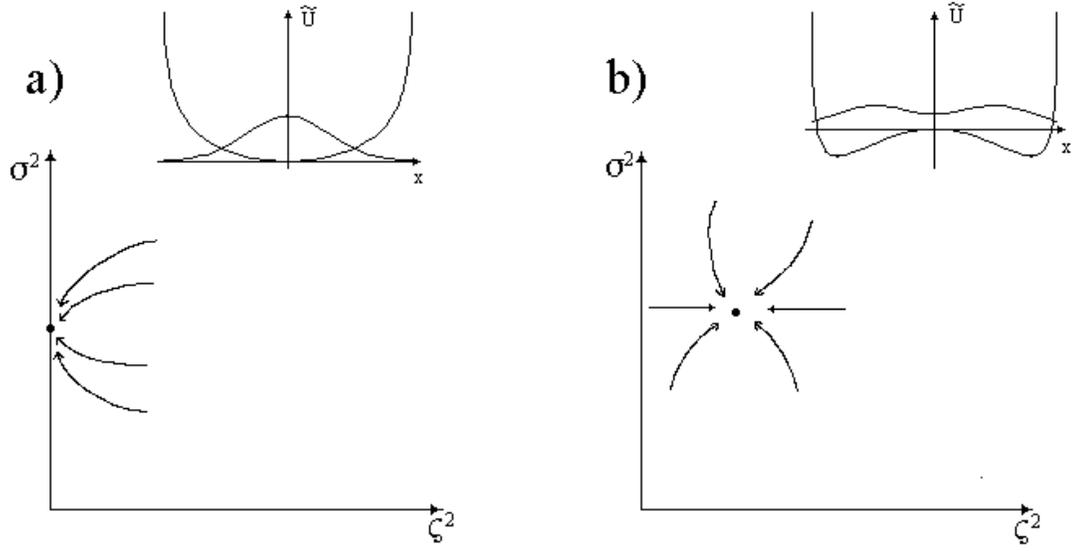}
\caption {Phase trajectories for the two-peak approximation, FENE model: {\bf a)} A stable
equilibrium on the vertical axis, one stable peak; {\bf b)} A stable equilibrium with $\varsigma>0$,
stable two-peak configuration.}
    \label{figFENE}
\end{centering}
\end{figure}

It is easy to check, that the formulas (\ref{projphi}) are indeed defining the projector: if
$f_{\sigma}$ (or $f_{\varsigma}$) is substituted there instead of the function $J$, then we will get
$\varphi_{\sigma}=1, \varphi_{\varsigma}=0$ (or $\varphi_{\sigma}=0, \varphi_{\varsigma}=1$,
respectively). Let us substitute the right part of the initial kinetic equations (\ref{530}),
calculated in the point $\Psi(q)=\Psi(\{\sigma,\varsigma\},q)$ (see the equation (\ref{Anz})) in the
equation (\ref{projphi}) instead of $J$. We will get the closed system of equations on $\sigma^{2},
\varsigma^{2}$ in the neighborhood of the Gaussian manifold.

This system describes the dynamics of the distribution function $\Psi$. The distribution function is
represented as the half-sum of two Gaussian distributions with the averages of distribution
$\pm\varsigma$ and mean-square deviations $\sigma$. All integrals in the right-hand part of
(\ref{projphi}) are possible to calculate analytically.

Basis $(f_{\sigma},f_{\varsigma})$ is convenient to use everywhere, except the points in the Gaussian
manifold, $\varsigma=0$, because if $\varsigma\rightarrow0$, then
\begin{eqnarray*}
f_{\sigma}-f_{\varsigma}=O\left(\frac{\varsigma^{2}}{\sigma^{2}}\right)\rightarrow0.
\end{eqnarray*}

Let us analyze the stability of the Gaussian manifold to the ``dissociation" of the Gaussian peak in
two peaks (\ref{Anz}). To do this, it is necessary to find first nonzero term in the Taylor expansion
in $\varsigma^{2}$ of the right-hand side of the second equation in the system (\ref{projphi}). The
denominator has the order of $\varsigma^{4}$, the numerator has, as it is easy to see, the order not
less, than $\varsigma^{6}$ (because the Gaussian manifold is invariant with respect to the initial
system).

With the accuracy up to $\varsigma^{4}$:
\begin{equation}\label{itog}
\frac{1}{\sigma^{2}}\frac{d\varsigma^{2}}{dt}=2\alpha\frac{\varsigma^{2}}{\sigma^{2}}+
o\left(\frac{\varsigma^{4}}{\sigma^{4}}\right),
\end{equation}
where $$\alpha=\kappa - \left. \frac{d U(M_{2})}{dM_{2}}\right|_{M_{2}=\sigma^2}.$$

So, if $\alpha>0$, then $\varsigma^{2}$ grows exponentially ($\varsigma\sim e^{\alpha t}$) and the
Gaussian manifold is unstable; if $\alpha<0$, then $\varsigma^{2}$ decreases exponentially and the
Gaussian manifold is stable.

Near the vertical axis $d\sigma^2/dt = 1+2\alpha \sigma^2$. The form of the phase trajectories is
shown qualitative on Fig. \ref{figFENEP}. Note that this result completely agrees with equation
(\ref{result1}).

For the real equation FPE (for example, with the FENE potential) the motion in presence of the flow
can be represented as the motion in the effective potential well $\tilde{U}(q)=U(q)-\frac{1}{2}\kappa
q^{2}$. Different variants of the phase portrait for the FENE potential are present on Fig.
\ref{figFENE}. Instability and dissociation of the unimodal distribution functions (``peaks") for the
FPE is the general effect when the flow is present.

The instability occurs when the matrix $\partial^{2}\tilde{U}/\partial q_{i}\partial q_{j}$ starts to
have negative eigenvalues ($\tilde{U}$ is the effective potential energy, $\tilde{U}(q)=U(q)- {1
\over 2} \sum_{i,j}\kappa_{i,j}q_{i}q_{j}$).

\subsubsection{Polymodal polyhedron}

The stationary polymodal distribution for the Fokker-Planck equation corresponds to the persistence
of several local minima of the function $\tilde{U}(q)$. The multidimensional case is different from
one-dimensional because it has the huge amount of possible configurations. An attempt to describe
this picture quantitative meet the following obstacle: we do not know the potential $U$, on the other
hand, the effect of molecular individualism \cite{Chu,DeGenne,Chu2} seems to be universal in its
essence, without dependence of the qualitative picture on details of interactions. We should find a
mechanism that is as general, as the effect. The simplest dumbbell model which we have discussed in
previous subsection does not explain the effect, but it gives us a hint: the flow can violate the
stability of unimodal distribution. If we assume that the whole picture is hidden insight a
multidimensional Fokker-Planck equation for a large molecule in a flow, then we can use this hint in
such a way: when the flow strain grows there appears a sequence of bifurcations, and for each of them
a new unstable direction arises. For qualitative description of such a picture we can apply a
language of normal forms \cite{ArVarGZ1995-1998}, but with some modification.

The bifurcation  in dimension one with appearance of two point of minima from one point has the
simplest polynomial representation: $U(q, \alpha)= q^4+ \alpha q^2$. If $\alpha \geq 0$, then this
potential has one minimum, if $\alpha < 0$, then there are two points of minima. The normal form of
degenerated singularity is $U(q)=q^4$. Such polynomial forms as $q^4+ \alpha q^2$ are very simple,
but they have inconvenient asymptotic at $q \rightarrow \infty$. For our goals it is more appropriate
to use logarithms of convex combinations of Gaussian distributions instead of polynomials. It is the
same class of jets near the bifurcation, but with given quadratic asymptotic $q \rightarrow \infty$.
If one needs another class of asymptotic, it is possible just to change the choice of the basic peak.
All normal forms of the critical form of functions, and families of versal deformations  are well
investigated and known \cite{ArVarGZ1995-1998}.

\begin{figure}
\begin{centering}
\includegraphics[width=140mm, height=170mm]{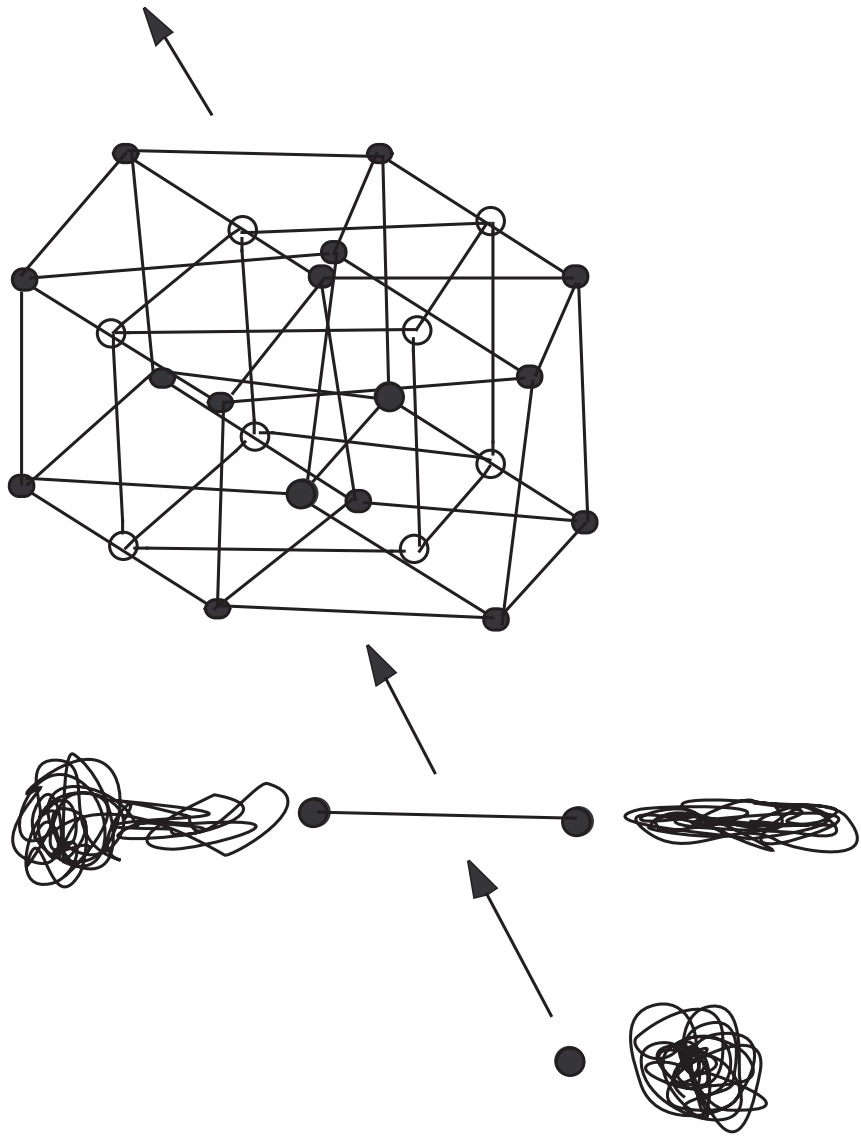}
\caption {Cartoon representing the steps of molecular individualism. Black dots are vertices of
Gaussian parallelepiped. Zero, one, and four-dimensional polyhedrons are drawn. Presented is also the
three-dimensional polyhedron used to draw the four-dimensional object. Each new dimension of the
polyhedron adds as soon as the corresponding bifurcation occurs. Quasi-stable polymeric conformations
are associated with each vertex. First bifurcation pertinent to the instability of a dumbbell model
in elongational flow is described in the text.}
    \label{Cartoon}
\end{centering}
\end{figure}

Let us represent the deformation of the probability distribution under the strain in multidimensional
case as a cascade of peak dissociation. The number of peaks will duplicate on the each step. The
possible cascade of peaks dissociation is presented qualitatively on Fig. \ref{Cartoon}. The
important property of this qualitative picture is the linear complexity of dynamical description with
exponential complexity of geometrical picture. Let $m$ be the number of bifurcation steps in the
cascade. Then
\begin{itemize}
\item{For description of parallelepiped  it is sufficient to describe
$m$ edges;}
\item{There are $2^{m-1}$ geometrically different conformations associated with $2^{m}$
vertex of parallelepiped (central symmetry halved this number).}
\end{itemize}
Another important property is the {\it threshold} nature of each dissociation: It appears in points
of stability loss for new directions, in these points the dimension of unstable direction increases.

Positions of peaks correspond to parallelepiped vertices. Different vertices in configuration space
present different geometric forms. So, it seems {\it plausible}\footnote{We can not {\it prove} it
now, and it is necessary to determine the status of proposed qualitative picture: it is much more
general than a specific model, it is the mechanism which acts in a wide class of models. The cascade
of instabilities can appear and, no doubt, it appears for the Fokker-Planck equation for a large
molecule in a flow. But it is not proven yet that the effects observed in well-known experiments have
exactly this mechanism. This proof requires quantitative verification of a specific model. And now we
talk not about a proven, but about the plausible mechanism which typically appears for systems with
instabilities.} that observed different forms (``dumbbels", ``half-dumbbels", ``kinked", ``folded"
and other, not classified forms) correspond to these vertices of parallelepiped. Each vertex is a
metastable state of a molecule and has its own basin of attraction. A molecule goes to the vertex
which depends strongly on details of initial conditions.

The simplest multidimensional dynamic model is the Fokker-Planck equation with quadratic mean field
potential. This is direct generalization of the FENE-P model: the quadratic potential $U(q)$ depends
on the tensor of second moments $\MM_{2}= \langle q_iq_j \rangle$ (here the angle brackets denote the
averaging). This dependence should provide the finite extensibility. This may be, for example, a
simple matrix generalization of the FENE-P energy: $$U(q)=\sum_{ij}K_{ij}q_iq_j, \: \KK=\KK^0 +
\phi(\MM_{2}/b), \: \langle U(q)\rangle ={\rm tr}(\KK \MM_{2}/b) $$ where $b$ is a constant (the
limit of extensibility), $\KK^0$ is a constant matrix, $\MM_{2}$ is the matrix of second moments, and
$\phi$ is a positive analytical monotone increasing function of one variable on the interval $(0,1)$,
$\phi(x)\rightarrow \infty$ for $x \rightarrow  1$ (for example, $\phi(x)=-\ln(1-x)/x$, or
$\phi(x)=(1-x)^{-1}$).

For quadratic multidimensional mean field models persists the qualitative picture of Fig.
\ref{figFENEP}: there is {\it non-stationary moleqular individualism for stationary ``molecular
collectivism"}. The stationary distribution  is the Gaussian distribution, and on the way to this
stationary point there exists an unstable region, where the distribution dissociates onto $2^m$ peaks
($m$ is the number of unstable degrees of freedom).

Dispersion of individual peak in unstable region increases too. This effect can deform the observed
situation: If some of the peaks have  significant intersection, then these peaks join into new
extended classes of observed molecules. The stochastic walk of molecules between connected peaks can
be observed as ``large non-periodical fluctuations". This walk can be unexpected fast, because it can
be effectively a {\it motion in a low-dimensional space}, for example, in one-dimensional space (in a
neighborhood of a part of one-dimensional skeleton of the polyhedron).

\subsection{Generalization: neurons and particles}

The Gaussian parallelepiped (\ref{parall}) seems to be a ``rigid" structure: the possibilities to
extend this ansatz, to make it more exact, but with preservation of more or less transparent
structure, are not obvious. The simple transformation can improve this situation. Let us mention the
obvious relation: $\exp(-(x-a)^2)=\exp(-x^2)\exp(2ax)\exp(-a^2).$ We can write the simple
generalization of equation (\ref{parall}):
\begin{equation}\label{neur}
\Psi(q)=\Psi^*(q)\prod_{i=1}^m \varphi_i((\varsigma_i,q)),
\end{equation}
where $\Psi^*(q)$ is the distribution density for one peak, for example, it may be the
multidimensional Gaussian distribution $\Psi^*(q)={1 \over(2\pi)^{n/2}\sqrt{\det
\Sigma}}\exp\left(-\frac{1}{2}\left(\Sigma^{-1}q, \: q \right)\right),$ $\varsigma_i$, ($i=1, \ldots,
m$) are vectors in the configuration space, $(\varsigma_i,q)$ is the usual scalar product,
$\varphi_i(x)$ are nonnegative functions of one variable $x$. for example, $\varphi_i(x)=A_i{\rm ch}
(\alpha_i x + \beta_i)$.

The form (\ref{neur}) is more flexible then original Gaussian parallelepiped (\ref{parall}). It gives
the possibility to extend the space for model adaptation. Functions of one variable $\varphi_i(x)$
are additional variables. They can form a finite-parametric family: For example,
$\varphi_i(x)=A_i{\rm ch} (\alpha_i x)$ give the Gaussian peaks, and if we use $\varphi_i(x)=A_i^+
\exp(\alpha_i^+ x) + A_i^- \exp(\alpha_i^- x)$, then we obtain a non-symmetric picture of shifted
peaks. On following steps we may use different spaces (or manifolds) of functions $\varphi_i(x)$ to
extend the approximation capacity of the ansatz (\ref{neur}).

Let us describe the tangent space $T$ for the ansatz (\ref{neur}) with functions $\lambda_i(x)=\ln
\varphi_i(x)$ from some space $L$.  The space of functions of $n$ variables $$L((\varsigma,
q))=\{\lambda((\varsigma, q))|\varphi \in L \}$$ corresponds to a given vector $\varsigma$ and the
space $L.$ The tangent space $T_{\Psi}$ for the ansatz (\ref{neur}) in a point $\Psi$ has a form:
\begin{equation}\label{TgNeu}
T_{\Psi}=\Psi \left[\sum_{i=1}^m L((\varsigma_i, q))+ \sum_{i=1}^m \left(\frac{d\lambda_i(x)
}{dx}\right)_{x=(\varsigma_i, q)}E^*
 \right],
\end{equation}
where $E^*$ is the space of linear functions of $q$.

If the space $L$ includes all sufficiently smooth functions, then to avoid intersection between
$L((\varsigma_i, q))$ and $\left(\frac{d\lambda_i(x)}{dx}\right)_{x=(\varsigma_i, q)}E^*$ it is
convenient to change in equation (\ref{TgNeu}) the space of all linear functions $E^*$ to the space
of linear functions orthogonal to $(\varsigma_i, q)$, $E^*_i=\{(\varsigma,q)|\varsigma \perp
\varsigma_i\}$ (without any change in the resulting space):
\begin{equation}\label{TgNeu'}
T_{\Psi}=\Psi \left[\sum_{i=1}^m L((\varsigma_i, q))+ \sum_{i=1}^m \left(\frac{d
\lambda_i(x)}{dx}\right)_{x=(\varsigma_i, q)}E^*_i
 \right].
\end{equation}
It means that for sufficiently rich spaces $L$ the vectors $\varsigma_i$ in the ansatz (\ref{neur})
could be chosen on the sphere, $(\varsigma_i,\varsigma_i)=1$, to provide the independence of
variables.

The form (\ref{neur}) appears as a quasiequilibrium distribution density in the following particular
case of the problem (\ref{maxent}):
\begin{eqnarray}\label{maxentneur}
S(\Psi)\rightarrow \max, \:  \int \delta(x-(\varsigma_i,q)) \Psi(q)d^n q=f_i(x), \: i=1,\ldots, m,
\end{eqnarray}
where $S(\Psi)$ is the Kullback-form Boltzmann-Gibbs-Shannon entropy which measures a deviation of
the distribution density $\Psi(q)$ from the equilibrium density $\Psi^*(q)$:
\begin{equation}\label{Kul}
S(\Psi)=-\int \Psi(q) \ln \left( \frac {\Psi(q)}{\Psi^*(q)} \right) d^n q.
\end{equation}
Hence, for fixed values of $\varsigma_i$ and for a space of arbitrary nonnegative smooth functions
$\varphi_i(x)$ the ansatz (\ref{neur}) is the quasiequilibrium approximation with macroscopic
variables $$ f_i(x)=\int \delta(x-(\varsigma_i,q)) \Psi(q)d^n q.$$

Let us define the ansatz manifold (\ref{neur}) as a union of the quasiequilibrium manifolds
(\ref{maxentneur}) for all sets of values $\{ \varsigma_i \}_{i=1}^m$ on the unit sphere. In this
case we can apply Proposition 4: The thermodynamic projector is the orthogonal projector on
$T_{\Psi}$ with respect to the entropic scalar product in the point $\Psi$: In the space of density
functions
\begin{equation}\label{espFPE}
\langle f|g\rangle_{\Psi}=\int \frac{f(q)g(q)}{\Psi(q)}dq,
\end{equation}
and in the conjugated space (for example, for functions $\lambda$ from space $L$ in
(\ref{TgNeu}),(\ref{TgNeu'}))
\begin{equation}\label{espFPEco}
\langle \mu| \lambda \rangle_{\Psi}^c=\int \mu(q)\lambda(q)\Psi(q)dq,
\end{equation}
where the scalar product for the conjugated space is marked by the upper index $c$.

The orthogonal projector $P$ on the direct sum of  subspaces
\begin{equation}\label{dirsum}
\sum_{i=1}^m L((\varsigma_i, q))+ \sum_{i=1}^m \left(\frac{d\lambda_i(x)}{dx}\right)_{x=(\varsigma_i,
q)}E^*_i
\end{equation}
is a sum of operators: $P=\sum_{i=1}^m (P_{\lambda_i}+P_{\varsigma_i})$, where
\begin{equation}\label{dirsumpro}
{\rm im}P_{\lambda_i}=L((\varsigma_i, q)), \: {\rm im} P_{\varsigma_i} = \left(\frac{d\lambda_i(x)
}{dx}\right)_{x=(\varsigma_i, q)}E^*_i.
\end{equation}
Operators $P_{\lambda_i}$, $P{\varsigma_i}$ can be founded from the definition of their images
(\ref{dirsumpro}) and the conditions: $P^2=P, \: P^+=P,$ where $P^+$ is the conjugated to $P$
operator with respect of the scalar product $\langle \: | \: \rangle^c$ (\ref{espFPEco}). From the
first equation of ($P^2=P$) it follows that each operator $A$ from the set $\QQ =
\{P_{\lambda_i}\}_{i=1}^m \bigcup \{P_{\varsigma_j}\}_{j=1}^m$ is a projector, $A^2=A$ (it may be not
orthogonal), and for any pair of distinct projectors  $ A,B \in \QQ$ the following inclusions hold: $
{\rm im}A \subseteq \ker B$, ${\rm im} B \subseteq \ker A$.

In a general case, the constructive realization of orthogonal projector requires solution of systems
of linear equations, or orthogonalization of systems of vectors, etc. We shall not discuss the
details of computational algorithms here, but one important possibility should be stressed. {\it The
orthogonal projection $P(J)$ can be computed by adaptive minimization of a quadratic form}:
\begin{equation}\label{adapta1}
\langle J-P(J) | J-P(J) \rangle_{\Psi} \rightarrow \min \: \mbox{for} \:  P(J) \in
T_{\Psi}
\end{equation}
The gradient methods for solution of the problem (\ref{adapta1}) are based on the following simple
observation: Let a subspace $L \subset E$ of the Hilbert space $E$ be the direct sum of subspaces
$L_i$: $L=\sum_i L_i$. The orthogonal projection of a vector $J \in E$ onto $L$ has an unique
representation in a form: $P(J)=x= \sum_i x_i,$ $x_i \in L_i$. The gradient of the quadratic form
$(J-x,J-x)$ in the space $L$ has the form:
\begin{equation}\label{gradad}
{\rm grad}_x(J-x,J-x)=2\sum_i P_i^{\perp}(J-x),
\end{equation}
where $P_i^{\perp}$ is the orthogonal projector on the space $L_i$. It means: if one has the
orthogonal projectors on the spaces $L_i$, then he can easy write the gradient method for the problem
(\ref{adapta1}).

The projected kinetic equations, $\dot{\Psi}=x$, $x \in T_{\Psi}$, with the equations for this
adaptive method, for example $\dot{x}=- h {\rm grad}_x\langle J-x | J-x \rangle_{\Psi}$, can be
solved together. For a rational choice of the step $h$ this system is stable, and has a Lyapunov
functional (for closed systems). This functional can be found as a linear combination of the entropy
and the minimized quadratic form $\langle J-x | J-x \rangle_{\Psi}$.

We consider the FPE of the form
\begin{equation}\label{FPE}
{\partial \Psi(q,t) \over \partial t} =\nabla_q \left\{D(q) \left[\Psi(q,t)(\nabla_q
U(q)-F_{ex}(q,t)) +\nabla_q \Psi(q,t)\right]\right\}.
\end{equation}
Here $\Psi(q,t)$ is the probability density over the configuration space $q$, at the time $t$, while
$U(q)$ and $D(q)$ are the potential and the positively semi-definite ($ (r,D(q)r)\ge 0$) symmetric
diffusion matrix, $F_{ex}(q,t)$ is an external force (we omit here such multipliers as $k_{\rm B}T$,
friction coefficients, etc). Another form of equation (\ref{FPE}) is:
\begin{equation}\label{FPEeq}
{\partial \Psi(q,t) \over \partial t} =\nabla_q \left\{D(q) \Psi^*(q) (\nabla_q - F_{ex}(q,t)) \left(
\frac {\Psi(q,t)}{\Psi^*(q)}\right)\right\},
\end{equation}
where $\Psi^*(q)$ is the equilibrium density: $\Psi^*(q,t)= \exp(-U(q))/ \int\exp(-U(p))dp.$ For the
ansatz (\ref{neur}) $\Psi(q,t)=\Psi^*(q)\exp \sum_i \lambda_i((\varsigma_i,q),t)$. For this ansatz
the left hand side of equation (\ref{FPEeq}) has the form
\begin{eqnarray}\label{J}
&&J(\Psi)=\Psi \left[\sum_i (\varsigma_i,D(q)\varsigma_i) \left(\frac{\partial^2 \lambda_i }{\partial
x^2}\right)_{x=(\varsigma_i,q)} + \sum_{i,j}(\varsigma_j,D(q)\varsigma_i) \left(\frac{\partial
\lambda_i}{\partial x}\right)_{x=(\varsigma_i,q)} \left(\frac{\partial \lambda_j}{\partial
x}\right)_{x=(\varsigma_j,q)} - \nonumber \right. \\ && \sum_i \left(\frac{\partial \lambda_i
}{\partial x}\right)_{x=(\varsigma_i,q)}((\varsigma_i,D(q)(\nabla_q U(q)+F_{ex}(q,t)))-(\nabla_q,
D(q)\varsigma_i))+ \nonumber \\ && \left. (\nabla_q U(q),D(q)F_{ex}(q,t))-(\nabla_q, D(q)F_{ex}(q,t))
\right],
\end{eqnarray}
where $\lambda_i = \lambda_i (x,t)$, and $(\:,\:)$ is the usual scalar product in the configuration
space.

The projected equations have the form:
\begin{equation}\label{proequneu}
\frac{\partial \lambda_i}{\partial t}=P_{\lambda_i}J(\Psi), \: \frac{d
\varsigma_i}{dt}=P_{\varsigma_i}J(\Psi),
\end{equation}
where the vector field $J(\Psi)$ is calculated by formula (\ref{J}), and the projectors
$P_{\lambda_i},$ $P_{\varsigma_i}$ are defined by equations (\ref{dirsumpro}). For adaptive methods
the right hand parts of equations (\ref{proequneu}) are solutions of auxiliary equations.

We can return from the ansatz (\ref{neur}) to the polymodal polyhedron: It corresponds to a
finite-dimensional multimodal approximation for each of equations (\ref{proequneu}). If the number of
maxima in the approximation of $\lambda_i(x)$ is $k_i$, then the number of peaks in the polymodal
polyhedron is $k=\prod_i k_i$.

For the further development of the approximation (\ref{neur}) it is possible add some usual moments
to the system (\ref{maxentneur}). These additional moments can include a stress tensor, and some
other polynomial moments. As a result of such an addition the equilibrium density in ansatz
(\ref{neur}) will be replaced to a more general nonconstant quasiequilibrium density.

The ansatz (\ref{neur}) can be discussed and studied from different points of view:
\begin{enumerate}
\item{It is a {\it uncorrelated particles} representation of kinetics:
The distribution density function (\ref{neur}) is a product of equilibrium density and one-particle
distributions, $\varphi_i$. Each particle has it's own one-dimensional configuration space with
coordinate $x=(\varsigma_i,q)$. The representation of uncorrelated particles is well known in
statistical physics, for example, the Vlasov equation is the projection of the Liouville equation
onto uncorrelated ansatz \cite{LPi}. There are three significant differences between the ansatz
(\ref{neur}) and usual uncorrelated ansatz: First, the ansatz (\ref{neur}) is not symmetric with
respect to particles permutation, second, the configuration spaces of particles for this ansatz are
dynamic variables. The third difference is: The ansatz (\ref{neur}) includes the equilibrium density
function explicitly, hence, the {\it uncorrelated particles} represent the  {\it nonequlibrium}
factor of distribution, and equilibrium correlations are taken into account completely.}
\item{It is a version of a {\it neural-network approximation} \cite{Arbib}. The components of the vector
$\varsigma_i$ are input synaptic weights for the $i$th neuron of the hidden layer, and $\ln
\varphi_i(x)$ is the activation function of this neuron. The activation function of the output neuron
is $\exp (x)$. There is no need in different input synaptic weights for the output neuron, because
possible activation functions of the neurons of the hidden layer form the linear space $L$, and any
real multiplier can be included into $\ln \varphi_i(x)$. The only difference from usual neural
networks is a relatively big space of activation functions on the hidden layer. Usually, the most
part of network abilities is hidden in the net of connections, and the only requirement to the
activation function is their nonlinearity, it is sufficient for the approximation omnipotence of
connectionism \cite{Cyb,GorA,GorWu}. Nevertheless, neural networks with relatively rich spaces of
activation functions are in use too \cite{Vecci,Mayer}.}

\end{enumerate}

\addcontentsline{toc}{section}{Conclusion: POET and the difference between ellipsoid and
parallelepiped}

\section*{Conclusion: POET and the difference between ellipsoid and
parallelepiped}

Let us introduce an abbreviation ``POET" (Projection-Of-Everything-Thermodynamic) for the
thermodynamic projector. POET transforms the arbitrary vector field equipped with the given Lyapunov
function into a vector field with the same Lyapunov function. It projects each term in kinetic
equations into the term with the same entropy production. Moreover, POET conserves the reciprocity
relations: if initial kinetics satisfies the Onsager relations, then the projected system satisfies
these relations too. Thus, the problem of persistence of thermodynamic properties in model reduction
is solved. POET is {\bf unique} operator which always  preserves the sign of dissipation, any other
important features of this operator follow from this preservation.

It is necessary to use POET even for reduction of kinetic models for open systems, because the
processes which produce the entropy in a closed system should produce the entropy in the open system
as well: The difference between open and closed systems is the presence of entropy outflow (or, what
is the same, of the free energy inflow), and the dissipative processes preserve their dissipativity.

One of the most important impacts of POET on the model reduction technology is the new possibility of
constructing thermodynamically consistent reduced model with almost arbitrary ansatz. On the other
hand, it gives the possibility to create thermodynamically consistent discretization of the problem
of model reduction \cite{Grids}.

The short memory approximation gives the coarse-grained equations. Entropy production for these
equations is larger than for initial equations. This approach allows us to produce systems with
dissipation from conservative systems, for example.

It should be stressed that all the constructions, equations and statements are valid for arbitrary
(linear or nonlinear) vector fields, classical or quantum, mechanical, or not. The only requirement
is: the projector field preserves the sign of dissipation, and such a field was constructed

In this paper we discussed the important example of ansatz: the multipeak models. Two examples of
these type of models demonstrated high efficiency during decades: the Tamm--Mott-Smith bimodal ansatz
for shock waves, and the the Langer--Bar-on--Miller approximation for spinodal decomposition.

The multimodal polyhedron appears every time as an appropriate approximation for distribution
functions for systems with instabilities. We create such an approximation for the Fokker--Planck
equation for polymer molecules in a flow. Distributions of this type are expected to appear in each
kinetic model with multidimensional instability as universally, as Gaussian distribution appears for
stable systems. This statement needs a clarification: everybody knows that the Gaussian distribution
is stable with respect to convolutions, and the appearance of this distribution is supported by
central limit theorem. Gaussian polyhedra form a stable class: convolution of two Gaussian polyhedra
is a Gaussian polyhedron, convolution of a Gaussian polyhedron with a Gaussian distribution is a
Gaussian polyhedron with the same number of vertices. On the other hand, a Gaussian distribution in a
potential well appears as an exponent of a quadratic form which represents the simplest stable
potential (a normal form of a nondegenerated critical point). Families of Gaussian parallelepipeds
appear as versal deformations with given asymptotic for systems with cascade of simplest
bifurcations.

The usual point of view is: The shape of the polymers in a flow is either a coiled ball, or a
stretched ellipsoid, and the Fokker--Planck equation describes the stretching from the ball to the
ellipsoid. It is not the whole truth, even for the FENE-P equation, as it was shown in ref.
\cite{IK00,Legendre}. The Fokker-Planck equation describes the shape of a probability cloud in the
space of conformations. In the flow with increasing strain this shape changes from the ball to the
ellipsoid, but, after some thresholds, this ellipsoid transforms into a multimodal distribution which
can be modeled  as the peak parallelepiped. The peaks describe the finite number of possible molecule
conformations. The number of this distinct conformations grows for a parallelepiped as $2^{m}$ with
the number $m$ of independent unstable direction. Each vertex has its own basin of attraction. A
molecule goes to the vertex which depends strongly on details of initial conditions.

These models pretend to be the kinetic basis for the theory of molecular individualism. The detailed
computations will be presented in following works, but some of the qualitative features of the models
are in agreement with some of  qualitative features of the picture observed in experiment
\cite{Chu,DeGenne,Chu2}: effect has the threshold character, different observed conformations depend
significantly on the initial conformation and orientation.

Some general questions remain open:

\begin{itemize}
\item{Of course, appearance of $2^m$ peaks in the Gaussian
parallelepiped is possible, but some of these peaks can join in following dynamics, hence the first
question is: what is the typical number of significantly different peaks for a $m-$dimensional
instability?}
\item{How can we decide what scenario is more realistic from the experimental
point of view: the proposed universal kinetic mechanism, or the scenario with long living metastable
states (for example, the relaxation of knoted molecules in the flow can give an other picture than
the relaxation of unknoted molecules)?}
\item{The analysis of random walk of molecules from peak to peak should be
done, and results of this analysis should be compared with observed large fluctuations.}
\end{itemize}

The systematic discussion of a difference between the Gaussian elipsoid (and its generalizations) and
the Gaussian multipeak polyhedron (and its generalizations) seems to be necessary. This polyhedron
appears generically as the effective ansatz for kinetic systems with instabilities.

{\bf Acknowledgements.} Dr. Patric Ilg calculated the curves for Fig. 1, M.S. Pavel Gorban calculated
the right hand sides of equations (\ref{projphi}) without Taylor expansion (with the same result).
Prof. Misha Gromov explained us some problems concerning asymptotic behaviour of number of maxima of
distribution functions (the vertices of polymodal polyhedron) in large dimensions. Professor H.C.
\"Ottinger asked fruitful questions about relations between model reduction and coarse-graining.
Comments of referees of Physica A helped us to improve the paper.

\end{document}